\documentstyle[12pt]{article}

\input epsf.tex
\input fps.sty

\newcommand{\be}{\begin{equation}}
\newcommand{\ee}{\end{equation}}
\newcommand{\bal}{\begin{array}{l}}
\newcommand{\eal}{\end{array}}
\newcommand{\bea}{\begin{eqnarray}}
\newcommand{\eea}{\end{eqnarray}}
\newcommand{\pat}{\partial}
\newcommand{\om}{\omega}

\newcommand{\xdot}{\dot{x}}

\newcommand{\half}{\frac{1}{2}}
\newcommand{\ptrue}{\phi_t}
\newcommand{\pfalse}{\phi_f}
\newcommand{\greapp}{\stackrel{>}{\sim}}
\newcommand{\lessapp}{\stackrel{<}{\sim}}
\newcommand{\vep}{\vec{p}}
\begin{document}
\baselineskip 16pt

\begin{titlepage}
\begin{flushright}
CALT-68-2050 \\
April 1996
\end{flushright}

\vskip 0.2truecm

\begin{center}
{\Large {\bf Tunneling in a Time Dependent Setting}}
\end{center}

\vskip 0.8cm

\begin{center}
{\bf Esko Keski-Vakkuri}$^1$ and {\bf Per Kraus}$^2$ 
\vskip 0.3cm
{\it California Institute of Technology \\
    Pasadena CA 91125, USA \\
    e-mail: esko or perkraus@theory.caltech.edu}
\end{center}

\vskip 2.2cm

\begin{center}
{\small {\bf Abstract: }}
\end{center}
\noindent
{\small A standard approach to analyzing tunneling processes in various 
physical contexts is to use instanton or imaginary time path techniques.
For systems in which the tunneling takes place in a time dependent setting,
the standard methods are often applicable only in special cases, {\em e.g.}
due to some additional symmetries. 
We consider a collection of time dependent tunneling problems to which
the standard methods cannot be applied directly, and present an algorithm,
based on the  WKB approximation combined with complex time path methods,
which can be used to calculate the relevant tunneling probabilities. This
collection of problems contains, among others, the
spontaneous nucleation of topological defects
in an expanding universe, the production of charged 
particle -- antiparticle pairs in a time dependent electric field, and
false vacuum decay in field theory from a coherently oscillating initial 
state.  To demonstrate the method, we present detailed calculations
of the time dependent decay rates for the last two examples. 
}
\rm 
\noindent
\vskip 3.0 cm

\small
\begin{flushleft}
$^1$ Work supported in part by a DOE grant DE-FG03-92-ER40701.\\
$^2$ Work supported in part by a DOE grant DE-FG03-92-ER40701 and by
a DuBridge Fellowship.
\end{flushleft}
\normalsize
\end{titlepage}

\newpage

\section{Introduction}

Problems involving quantum mechanical tunneling in a time dependent setting
can arise in
a wide variety of contexts, such as the ionization of atoms by strong laser
fields \cite{ERGEN},  
pair creation of charged particles in time dependent background
electromagnetic fields \cite{BI,MP,AUDRETSCH}, 
spontaneous nucleation of topological defects in
expanding universes \cite{BGV}, and false vacuum decay with  time 
dependent initial states or  time dependent potentials \cite{WIDROW}.
In some special cases, these systems can be
treated by standard instanton or imaginary time path methods; however, these
techniques have  limited applicability, and confusion often arises when
one tries to extend the analysis to more general time dependent situations.
For a discussion of various difficulties, see {\em e.g.} \cite{WIDROW}.

In this paper we will investigate a collection of generalized time 
dependent versions
of ``standard'' tunneling problems, where the textbook instanton and
imaginary time path methods are inapplicable due to the additional
time dependence. The models typically have Lagrangians with
an explicit time dependence arising from external backgrounds, or involve 
more complicated non-static initial states.
 Their unifying aspect is that they all
can be analyzed via a method that combines the use of the WKB approximation
 with  solutions of the classical equations of motion along
complex time paths. We will present a straightforward algorithm which can
be used to compute the relevant tunneling or nucleation rates for such 
systems. 

To give a concrete example of this method, let us consider  
pair creation  by a spatially constant electric field.  In order to identify
the specific
features associated with a time dependent field, it is useful to first 
review the simple case  of a static field.  This problem, first 
solved by Schwinger \cite{SCHWINGER}, is most
elegantly treated by an instanton approach.  Calling the state with no 
particles
present the false vacuum,  the decay rate is determined by the imaginary 
part of the false vacuum energy, which can be extracted from an imaginary 
time path integral over fields which approach the false vacuum 
at $\tau = -it = \pm \infty$.
In the semiclassical approximation one saturates the path integral by a 
solution to the (euclideanized)  equations of motion; the corresponding  
configuration is the instanton. The action of the instanton determines the
decay rate: $\Gamma \propto e^{-S_{\rm instanton}}$.

The instanton solution describes the nucleation of a particle, anti-particle 
pair in the background electric field. This can be seen directly by 
cutting the instanton in half. Half of the instanton solution corresponds to 
interpolating between the false vacuum at $\tau = -\infty$, and a turning 
point, which we can take to occur at $\tau=0$. The turning point 
configuration is that of a pair of particles momentarily at rest. If we 
were to continue evolving in imaginary time towards $\tau = \infty$, then the 
particles would  would converge and disappear, leaving the system in the 
false vacuum again. Instead, however, we can continue the solution to real 
time at the turning point, in which case the particles accelerate away from 
one another. Thus the full production process can conveniently be described 
by a combination of real and imaginary time evolution. 

For our purposes it is actually more convenient to consider the 
preceding discussion in the reverse order. We can start by 
considering the real time expanding  solution, and then 
consider evolving it back in time. Eventually we will 
reach the turning point, at which point we continue the 
evolution to imaginary time. If the particle separation 
proceeds to smoothly shrink to zero size in imaginary 
time, then the trajectory considered corresponds to a 
pair production process, and its action determines the 
decay rate. So, to summarize in a way that is most 
useful for the proceeding discussion, we look for 
expanding  solutions, which can be smoothly shrunk 
to zero size when evolved back along some complex time contour. 

Phrased in this way, it is apparent how to adapt the 
procedure to the more general problem of a time dependent 
electric field. We can again  look for solutions 
describing expanding pairs, but this time the continuation 
to complex time is more involved. Due to the time dependence 
in the problem, we no longer expect that the time contour 
along which the pair shrinks to zero size is one involving 
periods of purely real or purely imaginary time evolution; instead, the 
contour will be a more general curve in the complex time plane. Given 
that we can find such a contour, we can proceed to evaluate the 
action to determine the decay
 rate. The result will be a decay rate with non-trivial time dependence. 

By itself, the pair creation problem in a time dependent electric 
field has a long history. It was studied rigorously by Brezin and 
Itzykson \cite{BI}, using Schwinger's proper time approach. Marinov and 
Popov \cite{MP} treated it as a barrier penetration problem and
employed WKB methods. Further, as we discuss in the Appendix A, 
Audretsch \cite{AUDRETSCH} noticed that
the problem is isomorphic to overbarrier scattering in quantum mechanics. 
However, an advantage of the 
approach  outlined above is that it is easily adapted to other 
problems involving the decay of metastable states via the production of 
extended objects. Further, it yields an instantaneous pair production rate
with an explicit time dependence, so that one sees a time modulation in the 
flux of produced particles. The result for the
pair production rate in \cite{MP} applies only at specific times.  

 As an example of adapting our approach to other physical processes, one can
consider generalizing the computation 
of false vacuum decay in field theory, which proceeds through the 
nucleation of bubbles of true vacuum, to
include field potentials with explicit time dependence.  
A particularly interesting
source of time dependence arises from  expanding 
universes, where it is expected that the expansion gives rise to the 
spontaneous nucleation of monopoles, strings, and domain walls.  
The nucleation rate has been
computed for a very specific case, namely De Sitter space;  but 
this is not in fact
a time dependent problem, as the De Sitter geometry is static.  
For non-static
geometries the more general approach discussed above is required.  

In the next section we write down an action which is general enough to treat 
the various processes
we have referred to,  and then give an 
algorithm by which one can compute the time
dependent
nucleation rate of the corresponding objects. 
In most cases, several steps in the
procedure must be performed numerically.  
The simplest case, in which almost 
everything can be done analytically, is pair production in a 
time dependent
electric field.  
We perform these steps in section 3, showing that the production
rate takes a compact integral form.  A special case
of this formula was derived before in \cite{MP}. We review
the connection of the problem to the problem of above barrier scattering 
in quantum mechanics. This discussed in detail in Appendix A.  We then 
study the particular example of a sinusoidally varying field, and analyze
the instantaneous pair production rates.  
In section 4 we turn to the other source of time dependence mentioned 
above, arising  from the initial state rather than from external sources.  
Specifically, we
consider a field theory with a local, but not global, minimum, and take the 
initial
state to be one in which the field is undergoing coherent oscillations 
about the
local minimum.  We are  able to calculate analytically for small oscillations,
and to obtain the leading correction to the decay rate.  Some computational 
details
are relegated to Appendix B.  Finally, in section 5 we summarize our 
conclusions
and discuss directions for further study.  

\section{Time Dependent Tunneling}

In this section we discuss our general approach to tunneling in 
models with explicit
time dependence.  Our aim  is to show, using complex time contours, how 
such systems
can be treated by a natural extension of the standard instanton method.  
To start 
with, we assume that there is an underlying field theory description of the 
model
under consideration, and that that the system is initially  in some m
etastable state.
We further assume  that the state can decay via quantum tunneling, and that 
the decay
occurs through the production of objects which can be described by a 
first quantized
action.  A wide variety of such objects can be described by the 
following action:
\be
S~=~\int  dt [-a(x,t)\sqrt{1-\dot{x}^2}~+~b(x,t)] .
\label{genS}
\ee
Some specific cases of interest are:
\begin{itemize}
\item $a(x,t)~=~m {\mbox \ \ \ \ ; \ \ \ \ } b(x,t)~=~qE(t)x$
\end{itemize}
This is the relativistic  action of a particle of 
mass $m$ and charge $q$ moving  in 
a time dependent electric field $E(t)$.  As will be 
discussed in detail, this is the
appropriate action for considering pair production 
due to the electric field.
\begin{itemize}
\item $a(x,t)~=~4\pi\sigma(t)x^2 {\mbox\ \ \ \ ; \ \ \ \ } 
  b(x,t)~=~\frac{4}{3}\pi\rho(t)x^3$
\end{itemize}
This is the action of a spherical ``bubble'' of radius $x$, with 
time dependent 
surface tension $\sigma(t)$ and bulk energy density $\rho(t)$.  
It describes false 
vacuum decay in a field theory from a time dependent 
initial state or in a
time dependent potential, in instances  where
the thin wall approximation is valid.
\begin{itemize}
\item $a(x,t)~=~mc(t) {\mbox\ \ \ \ ; \ \ \ \ }  b(x,t)~=~0$
\end{itemize}
This is the action of a massive particle moving in the metric 
$$
ds^2~=~c^2(t)[dt^2-dx^2-s^2(x)(d\theta^2+{\sin{\theta}}^2d\phi^2)]
$$
at fixed $\theta$ , $\phi$.  Here the choices $s(x)=\sin{x},x,\sinh{x}$ give 
closed, flat, and open Robertson-Walker universes.
\begin{itemize}
\item $a(x,t)~=~2\pi\mu c^2(t)s(x) {\mbox \ \ \ \ ; \ \ \ \ }  b(x,t)~=~0$
\end{itemize}
We then have the action of a circular cosmic string 
moving in the above metric.  The
string is located at $\theta=\pi/2$ and is centered at $x=0$.
\begin{itemize}
\item $a(x,t)~=~4\pi\sigma c^3(t)s^2(x){\mbox \ \ \ \ ; \ \ \ \ }  b(x,t)~=~0$
\end{itemize}
This, similarly, describes a spherical domain wall in the above metric, again 
centered at $x=0$.

One might question the applicability of first quantized action to 
describe these
systems, which are fundamentally field theories.  In the electric 
field example, it
is possible to make the connection rigorous and explicit, as is 
discussed 
in Appendix A.  We can see no reason why the connection should not be 
valid in the
other examples as well.

We now explain how the action (\ref{genS}) can be used to 
compute the spontaneous
creation rate of the objects which it describes.  To begin, we should find the
classical, real time, trajectories.  The equations of motion are
\be
\frac{d}{dt}\left[\frac{a(x,t)\dot{x}}{\sqrt{1-{\dot{x}}^2}}\right]~=~
-\sqrt{1-{\dot{x}}^2} a'(x,t)+b'(x,t) \,
\label{eqmot}
\ee
where $'$ denotes $\partial/\partial x$.  For the purposes of tunneling, the 
relevant trajectories are those which emanate from a turning point.  
For a trajectory
$x(t)$, the existence of a turning point at time  $t_f$ means that the 
canonical 
momentum vanishes there:
$$
\left.  p(t_f)~=~\frac{a(x,t)\dot{x}}{\sqrt{1-{\dot{x}}^2}}
                 \right|_{t=t_f}~=~0 .
$$
In most cases, this condition will be simply $\dot{x}=0$.  A nucleation process
corresponds to a trajectory which smoothly shrinks the object down to 
zero size when
evolved back in time along a complex time contour. The continuation from 
real to
complex time occurs at the turning point.  Shrinking to zero size  
means that $x=0$
\footnote{In the electric field case, $x$ denotes the position 
of a particle whose
anti-particle is located at $-x$. $x=0$ thus corresponds to zero separation.}
in the flat space examples, whereas in curved space it is the physical 
size $c(t)s(x)$
which is required to go to zero.  The condition that the shrinking to 
zero size be
smooth is most easily seen in the flat space examples.  Then we 
require that 
$\dot{x}\rightarrow 0$ when $x \rightarrow 0$; otherwise in the 
electric field case,
for instance, the joining of the particle trajectory, $x(t)$, and the 
anti-particle
trajectory, $-x(t)$, will be singular.  In curved space a slightly 
more detailed
analysis is necessary, depending on the specific form of $c(t)$.  
To summarize, the
trajectories of interest 
satisfy $p(t_f)=0$  , $x(t_0)=0$ , $\dot{x}(t_0) \rightarrow
 \infty$, $t_f=$ real, $t_0=$ complex , $x(t)=$ real.  The problem is 
then: given
some time $t_f$, find an initial size $x(t_f)$ and complex 
time $t_0$ such that the
above conditions are satisfied.  

Since $x$ is required to be real while $t$ is complex, the easiest 
way to proceed 
is to rewrite the equation of motion (\ref{eqmot}) as an equation for $t(x)$:
\be
\frac{d}{dx}\left[\frac{a(x,t)}{\sqrt{t'^2-1}}\right]~=~-a'(x,t)\sqrt{t'^2-1}
+b'(x,t)t' .
\label{eqt}
\ee
The conditions on $t(x)$ become
$$
\left. p(t_f)~=~\frac{a(x,t)}{\sqrt{t'^2-1}}\right|_{t_f}~=~0 
        {\mbox \ \ \ \ ; \ \ \ \ } t'(0)~=~0.
$$
The advantage of this form is that it is straightforward to solve (\ref{eqt})
numerically, even if it is not possible to do so analytically.  
Then we can search
for an initial coordinate value $x_f=x(t_f)$ which 
leads to  $t'(0)=0$.  Having found an
appropriate tunneling trajectory, $t(x)$, we can proceed to 
evaluate its action.
Again, it is easiest to change variables in (\ref{genS}), yielding
$$
S~=~\int_{0}^{x_f} dx \, [-a(x,t(x))\sqrt{t'^2(x)-1}+b(x,t(x))t'(x)] .
$$
The limits of integration run from $0$ to $x_f$ simply because that 
gives the entire
contribution to the imaginary part of the action.  The imaginary 
part of the action
determines the decay rate:
$$
\Gamma(t_f)~=~e^{-2{\rm Im }[S(t_f)]} .
$$
Notice that the decay rate depends on the time $t_f$ which is the end point
of the complex time path. Therefore, the result 
really is an {\em instantaneous} decay rate: the decay probability per unit
time (per unit volume) at a time $t_f$. Naturally, this is an idealized
situation, a more realistic decay rate would involve a time average depending
on the characteristic time scales and physical limitations of the given 
situation. However, even after such averaging, the decay rate characterizing
the production rate  will typically still be time modulated,
and the time dependence can lead to observable consequences.

\section{Particle Creation in a Time Dependent Electric Field} 

As a concrete illustration of the method discussed in the previous 
section, we will
compute the probability for production of charged particles 
in a time dependent
electric field. As noted previously, a rigorous 
approach to this problem involves
starting from the second quantized field theory. For instance, \cite{BI}
analyzed the pair production in an alternating electric field through a
Schwinger proper time approach. An alternative calculation could proceed
through a Bogoliubov transformation relating the `in' and `out' asymptotic 
states;
this   approach is outlined in Appendix A. However, as we 
also discuss in Appendix A, the same results can be 
obtained from the more intuitive 
first quantized approach which we consider in this section.  

We begin by considering the simple case of a static field. That a 
(spatially and temporally) constant electric field should be capable of 
creating charged pairs is evident on energetic grounds, since the 
requisite energy, $2mc^2$, needed to create the pair is supplied 
by the electric field, if
we separate the particles by a distance $2mc^2 / qE$. The creation 
rate is proportional to the probability of the particles, initially 
located at the same point, to tunnel to this separation.  
Our starting point is the spin-0 point particle action in the 
presence of a constant
electric field:
\be
 S = - \int \ dt [m\sqrt{1-\dot{x}^2} - qEx] \ .
\label{action1}
\ee
For simplicity, we have taken space to be one dimensional. For ease of 
comparison
with the discussion in Appendix A, it is actually simpler to integrate 
by parts and
use the action
\be
S = - \int \ dt [m\sqrt{1-\dot{x}^2} + qEt\dot{x}] \ .
\label{action2}
\ee
(This amounts to a different gauge choice for the gauge potential $A_{\mu}$.)
The equation of motion is
$$
     m \frac{d}{dt} \left[ \frac{\xdot}{\sqrt{1 - \xdot^2}} \right] = q E \ \ .
$$
The solution is 
$$
   x(t) = x(t_0) + \frac{m}{qE} \left\{ \sqrt{1 + \left[ \frac{qE}{m} 
            (t-t_0) + \frac{\xdot (t_0)}{\sqrt{1 - \xdot^2 (t_0)}} \right]^2}
             - \frac{1}{\sqrt{1-\xdot^2 (t_0)}} \right\} \ \ .
$$
Since we are interested in a tunneling process, we continue 
these trajectories to imaginary time $t \mapsto -i\tau$. Let us 
consider the particular trajectory for 
which $\tau_0 = -m/qE; \ x(\tau_0)=0;\ \xdot (\tau_0) = i\infty$.
This yields the circle
$$
      x^2 (\tau) + \tau^2 = (\frac{m}{qE})^2 \  \ ,
$$
a quarter of it is shown in Figure 1.

\begin{center}
\leavevmode
\fpsxsize 2in
\fpsbox{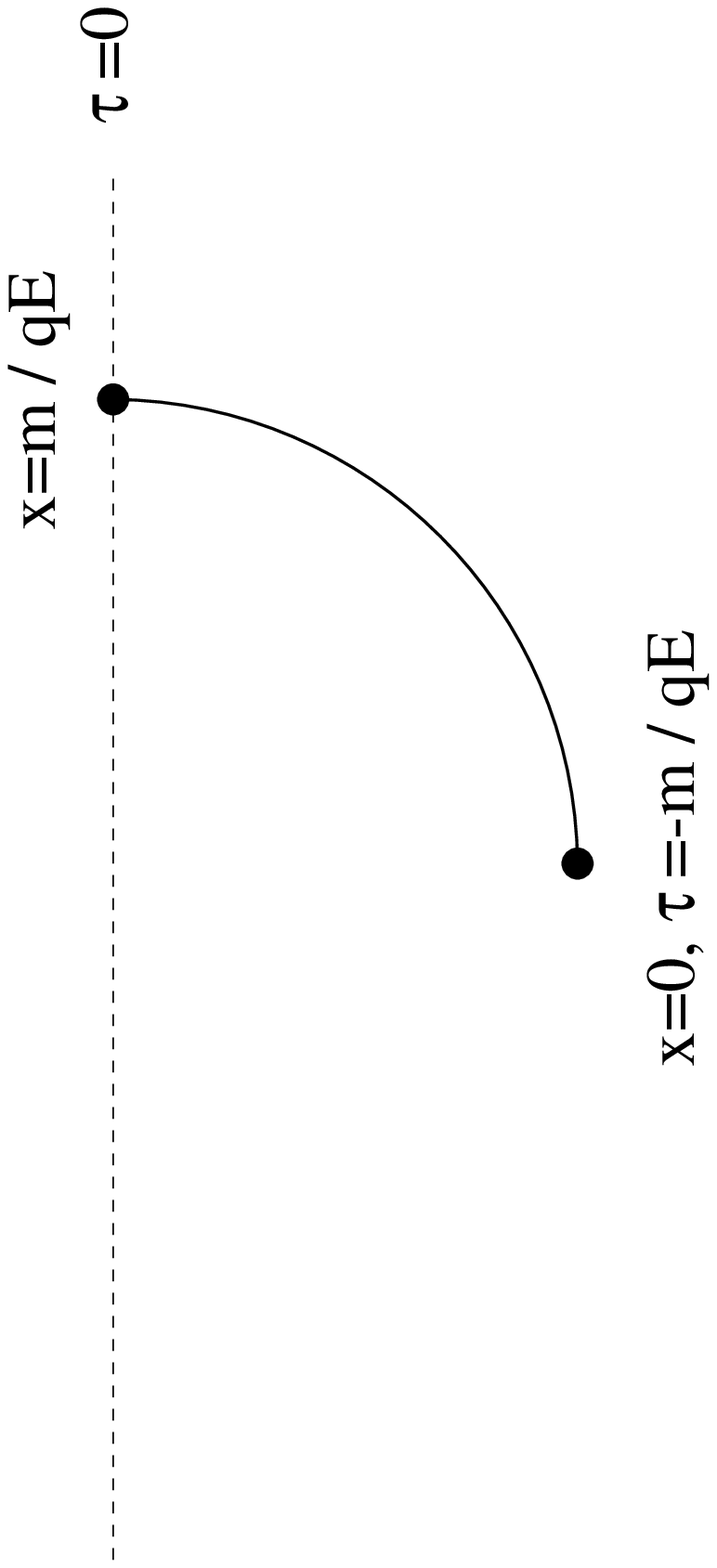}
\end{center}
\vskip 0.1 true cm
\begin{quotation}
\small
\noindent
{\bf Figure 1:}
A trajectory for half of the tunneling process.
\end{quotation}

The particle comes to rest at $\tau =0$ after having traveled a distance
$\Delta x = m / qE$. This portion of the path (one quarter of the circle) 
describes half of the tunneling process. The action for this 
portion is found by substituting the trajectory $x(\tau )$ into 
$$
     S = i \int^0_{-m/qE} \ d\tau [m \sqrt{1 + (\frac{dx}{d\tau})^2} 
                                    + qE\tau \dot{x}(\tau )] \ \ ,
$$
where $\dot{}$ now means $d/d\tau$.
This yields
$$
        S = \frac{i\pi m^2}{4qE} \ \ .
$$
The full tunneling process is described by a semicircle (see Figure 2.)

\begin{center}
\leavevmode
\fpsxsize 1.5in
\fpsysize 4in
\fpsbox{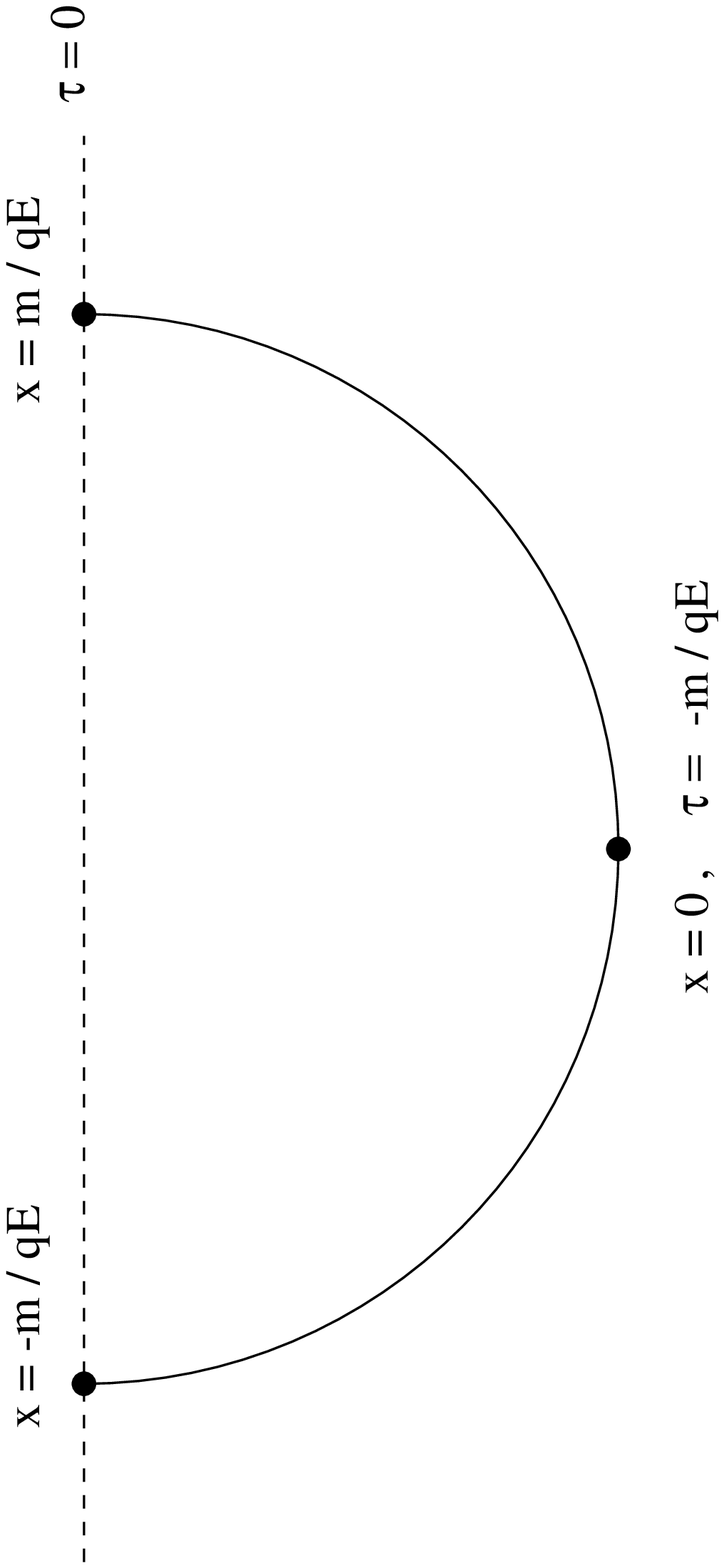}
\end{center}
\vskip 0.1 true cm
\begin{quotation}
\small
\noindent
{\bf Figure 2:}
A trajectory for the full tunneling process.
\end{quotation}

Whereas the part of the trajectory between $x=0$ and $x=m/qE$ depicts
the creation of a particle with charge $q$, the part between $x=0$ and
$x = -m/qE$ depicts the creation of the corresponding antiparticle of 
charge $-q$. We see that the antiparticle trajectory is 
obtained from the particle 
trajectory by $q \mapsto -q; \ x(\tau ) \mapsto -x(\tau )$. It is easy to 
see that this path yields a solution for the equation of motion and has the
same action as the particle trajectory. The total tunneling action is thus
$$
        S_{\rm total} = \frac{i \pi m^2}{2 q E} \ \ .
$$
The rate of pair creation is found by squaring the tunneling amplitude:
$$
       \Gamma = \left| e^{-{\rm Im}[S_{\rm total}]} \right|^2 
              = e^{-\frac{\pi m^2}{qE}} \ \ .
$$
The exact result obtained by Schwinger \cite{SCHWINGER} is
$$
 \Gamma = \frac{(qE)^2}{(2\pi )^3} \sum^{\infty}_{n=1} \ \frac{(-1)^{n+1}}{n^2}
              e^{-\frac{n\pi m^2}{qE}} \ \ .
$$
The WKB result gives a good approximation when $\pi m^2 / qE \gg 1$.  
Finally, we note
that
after being created, the particles move along hyperbolas, as is 
seen by continuing 
the trajectories back to real time. 

We now generalize the analysis to treat a time dependent electric field. 
As discussed in the previous section,
 we expect that this can be accomplished by finding a complex time path 
connecting the 
initial and final positions. 
\newpage
In particular, we will look for a path as shown in Figure 3.

\begin{center}
\leavevmode
\fpsxsize 2in
\fpsbox{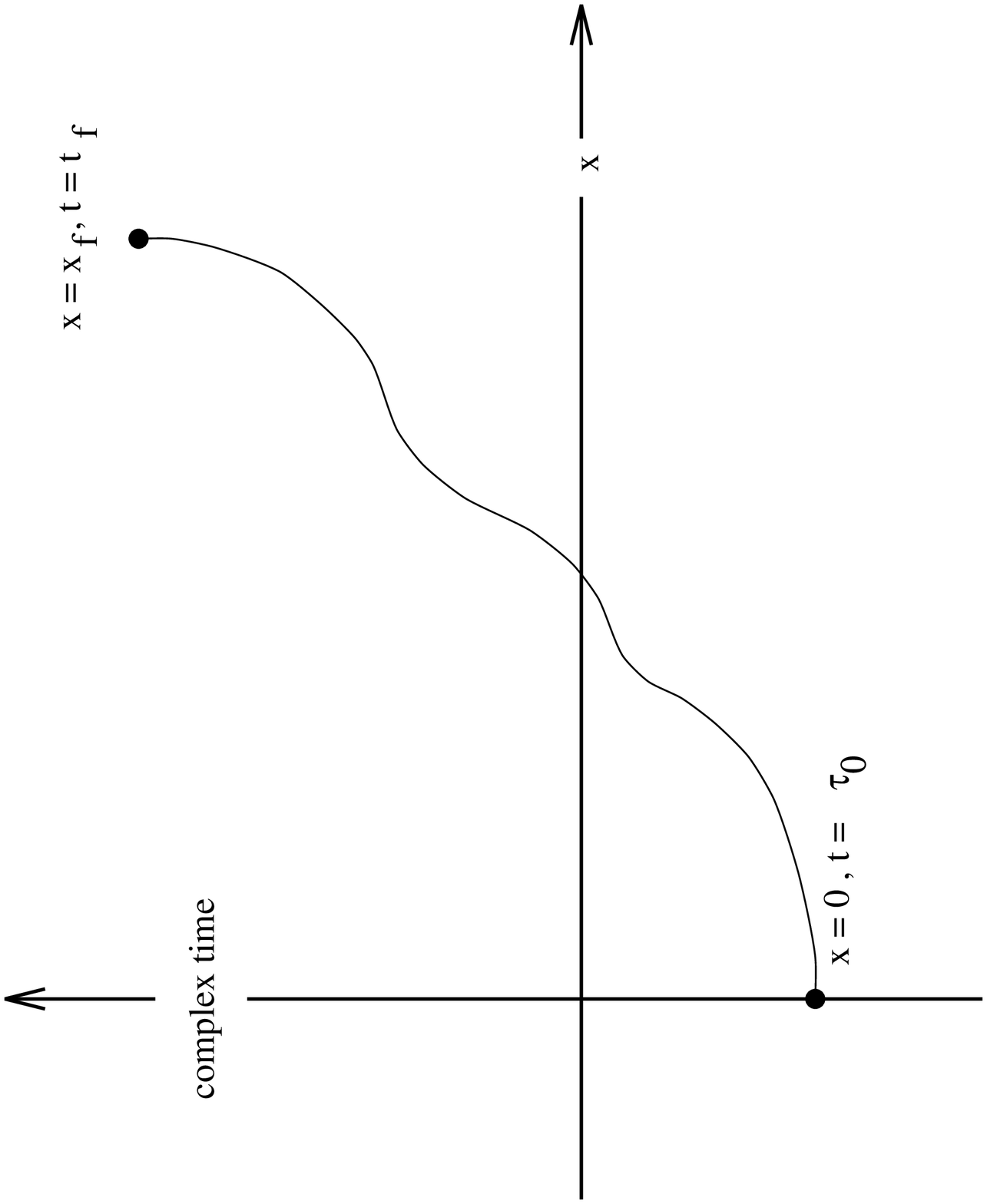}
\end{center}
\vskip 0.1 true cm
\begin{quotation}
\small
\noindent
{\bf Figure 3:}
A trajectory for half of the generic tunneling process.
\end{quotation}

\noindent
which describes the creation of a particle at time $t=t_f$. The vertical axis
no longer represents purely real or imaginary time, but rather some 
more general complex time direction (such that $t_f$ is real). 
At $t=t_f$ the particle is at rest, $\xdot (t_f) =0$.
At $t=\tau_0$ the velocity should be singular, $\xdot (\tau_0) \rightarrow 
\infty$, so that the trajectory for the antiparticle can be smoothly 
joined on to the particle trajectory. Both of these conditions 
were of course satisfied in the static field case. The remarkable 
aspect of this problem, as we shall see, is that these
conditions allow us to determine the tunneling action without 
having to find the complex time path explicitly.

The action, after integrating by parts, has the form
\be
S = - \int \ dt [m\sqrt{1-\dot{x}^2} - qA(t)\dot{x}] \ .
\label{action3}
\ee
where we have defined 
\be
A(t)=-\int_{t_f}^{t} dt' \,E(t').
\label{defA}
\ee
  The equations of motion then yield
$$
 \xdot (t) = -\frac{qA(t)/m}{\sqrt{1 + q^2A^2(t)/m^2}} \ \ .
$$
We have set the conserved momentum
$p_x = \pat L / \pat \dot{x}$ to zero, so that
$\xdot$ respects the condition $\xdot (t_f) =0$. Integrating:
$$
 x(t) = -\frac{q}{m}\int^t_{t_f} \ dt' 
              \frac{A(t')}{\sqrt{1 + q^2A^2 (t')/m^2}} + x(t_f) \ \ .
$$
Now, $t_0$ is determined by requiring $\xdot (t_0) \rightarrow \infty$. For
a nonsingular $E(t)$, this implies $A(t_0)=\pm im/q$. We choose to 
set $A(t_0) = -im/q$.

We will now write the action in terms of $A(t)$ and see that it takes a 
simple form. Substituting  the 
expression for $\dot{x}$  into (\ref{action3}) gives
$$
     S = -m \int^{t_f}_{t_0} \ dt \sqrt{1 + q^2A^2 (t)/m^2} \ \ .
$$
Now we change the integration variable from $t$ to $v=-qA/m$ using
$$
        dt = \frac{m\, dv}{ qE(t(v))} \ \ ,
$$
where $t(v)$ is found by inverting (\ref{defA}). Then,
$$
      S = -\frac{m^2}{q} \int^0_i \ dv \frac{\sqrt{1+v^2}}{E(t(v),t_f)}
        = \frac{im^2}{q} \int^{\pi /2}_0 \ d\theta \frac{\cos^2 \theta}
           {E(t(i\sin \theta ),t_f)} \ \ .
$$
This is the action corresponding to the creation of the charge $q$ particle. 
As before, the action for the antiparticle is obtained by replacing 
$q \mapsto -q; \ x(t) \mapsto -x(t)$, which yields the same result as for 
the particle. Therefore, the pair creation rate is given by
$$
  \Gamma (t_f) = \exp \{ -4 {\rm Im} [S(t_f)] \} \ \ .
$$
To recapitulate, the essential trick that was used was to map the potentially
complicated complex time contour into the complex $v$-plane, where it always
takes the simple form of a line from 0 to i. This simplifies the problem 
considerably, since it is no longer necessary to try to find the complex
time contour. We only need to know its image 
in the $v$-plane, and we do.

\subsection{Example of Time Dependent Pair Creation}

Having obtained the general result\footnote{In this subsection, we restore
$\hbar$ and $c$.}
\be
      S = \frac{im^2 c^3}{q} \int^{\pi /2}_0 \ d\theta \frac{\cos^2 \theta}
           {E(t(i c\sin \theta),t_f)} 
\label{generalS}
\ee
for the action along the complex time path, we shall now evaluate the
pair creation rate
$$
     \Gamma (t_f )= \exp \{ -\frac{4}{\hbar} {\rm Im} [S(t_f)] \}
$$
in an example case of a time dependent electric field. 
But first, we check the formula with a constant electric 
field $E = E_0$. Now the denominator in (\ref{generalS}) is a constant,
and what remains is an elementary integral, so we easily obtain the standard
WKB result for the pair creation rate
\be
      \Gamma = \exp \{ -\frac{\pi m^2 c^3}{\hbar q E_0} \} \ \ .
\label{standardG}
\ee
This pair creation rate is generally truly small. Even for strongest electric
fields obtained in a laboratory, $E_0 \sim 10^{11}$ N/C \cite{strongE},
the exponent is still enormous: using
$q=e\sim 10^{-19} $ C, $mc^2 \sim 
m_ec^2 \sim 10^6$ eV $\sim 10^{-13}$ Nm, $c \sim 10^8$ m/s, $\hbar 
\sim 10^{-34} $ Js, we get a vanishingly small production rate
$$
    \Gamma \sim \exp \{ -10^8 \} \\ .
$$

Now, let us consider an example of a time dependent electric field, an
oscillating strong electric field
$$
 E(t_f) = E_0 \cos \om t_f \ .
$$
Note that no magnetic field is required to satisfy the Maxwell's equations,
only an alternating uniform current density. This case is 
different from pair creation in the background of an 
electromagnetic plane wave which
was considered originally by Schwinger \cite{SCHWINGER} in the sense
that the intensity of the plane wave is time independent.

We first find $v$ and $E(t(v),t_f)$:
$$
  v(t,t_f) = \frac{q}{m}\int_{t_f}^{t}dt'\,E(t')~=~
\frac{qE_0}{m\om} (\sin \om t - \sin \om t_f)
$$
and
$$
  E( t(ic\sin \theta),t_f) = E_0 \sqrt{1 - (\frac{im\om c\sin \theta}
               {qE_0} + \sin \om t_f)^2} \ \ .
$$
After changing the integration variable to $x = \sin \theta$, the action
can be written as
$$
  S = \frac{im^2c^3}{qE_0} \int^1_0 \ dx \frac{\sqrt{1-x^2}}
         {\sqrt{1 - (i\frac{m\om c}{qE_0}x + \sin \om t_f )^2} } \ \ .
$$
Finally, isolating the imaginary part of the action we find the instantaneous
pair creation rate to be of the form
$$
\Gamma = \exp \left\{ -\frac{\pi m^2 c^3}{\hbar q E_0} \ I(t_f) \right\} 
$$
where $I(t_f)$ is a  modulation factor which characterizes the time dependence
 (compare
with (\ref{standardG})). We find that
\begin{eqnarray*}
 I(t_f) &=& \frac{2\sqrt{2}}{\pi} \int^1_0 \ dx 
\frac{\sqrt{1-x^2}\sqrt{r(x,t_f) + u(x,t_f)}}{r(x,t_f)} \ \ , \\ \nonumber
u(x,t_f) &=& 1 + a^2x^2 - \sin^2 \om t_f \\ \nonumber
r(x,t_f) &=& \sqrt{u^2(x,t_f)+ 4 a^2 x^2 \sin^2 \om t_f} \\ \nonumber
a &\equiv& \frac{m\om c}{qE_0} \ \ .
\end{eqnarray*}
This integral can be evaluated numerically, we show a plot of
$I(t_f)$ for $a=1$ ($\om = qE_0 / mc$):

\newpage

\begin{center}
\leavevmode
\fpsxsize 3in
\fpsbox{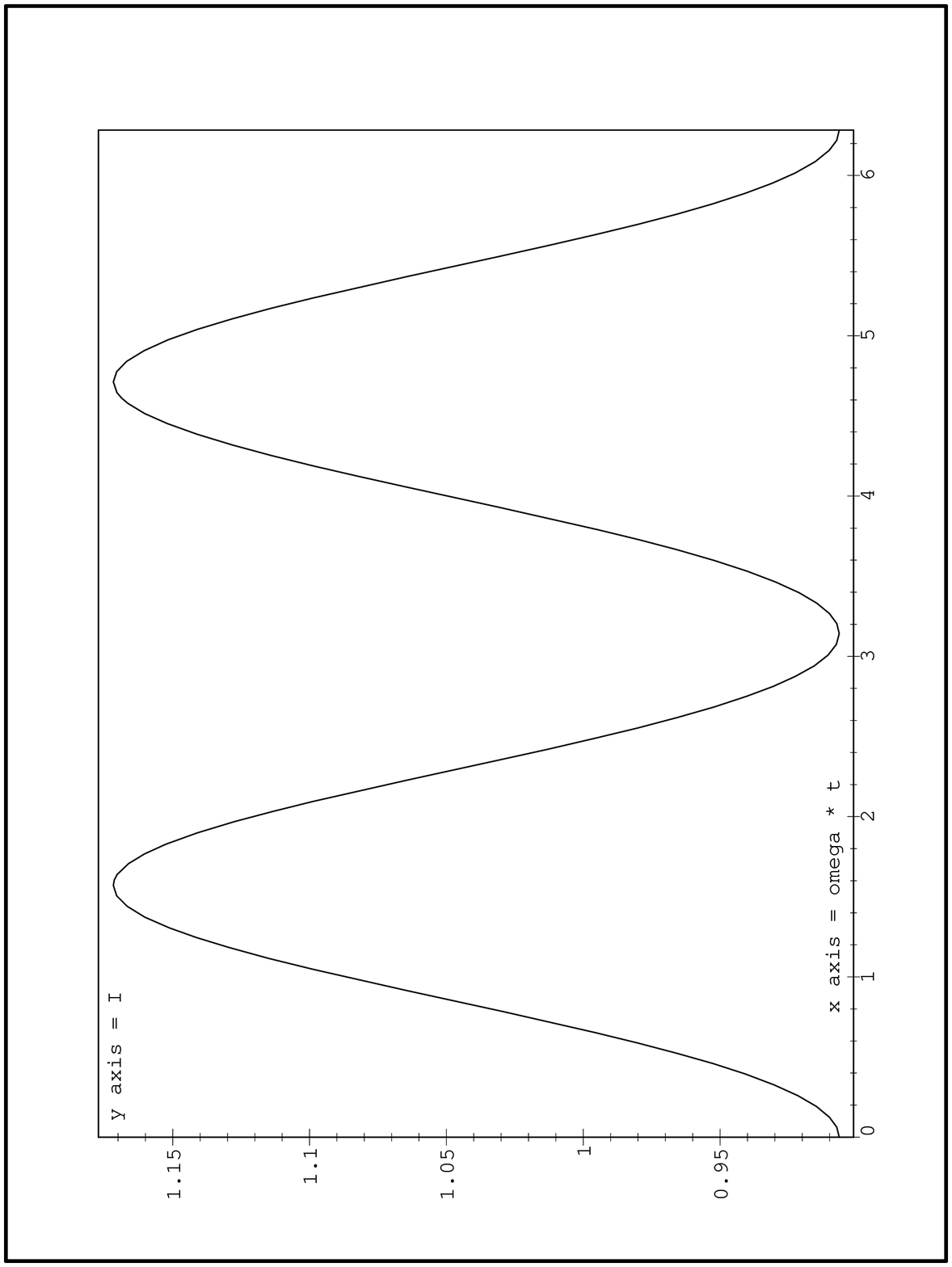}
\end{center}
\vskip 0.1 true cm
\begin{quotation}
\small
\noindent
{\bf Figure 4:}
An example plot of the time modulation function $I(t_f)$ in the action.
\end{quotation}

We can see that $I(t_f)$ is a periodic function of time. It reaches a
minimum value at times $\om t_f = 0 + n\pi$ and a maximum at
$\om t_f = \pi / 2 + n\pi$. The dependence of the minimum and maximum values
of $I$ on the frequency $\om$ turns out to be interesting:

\begin{center}
\leavevmode
\fpsxsize 3in
\fpsbox{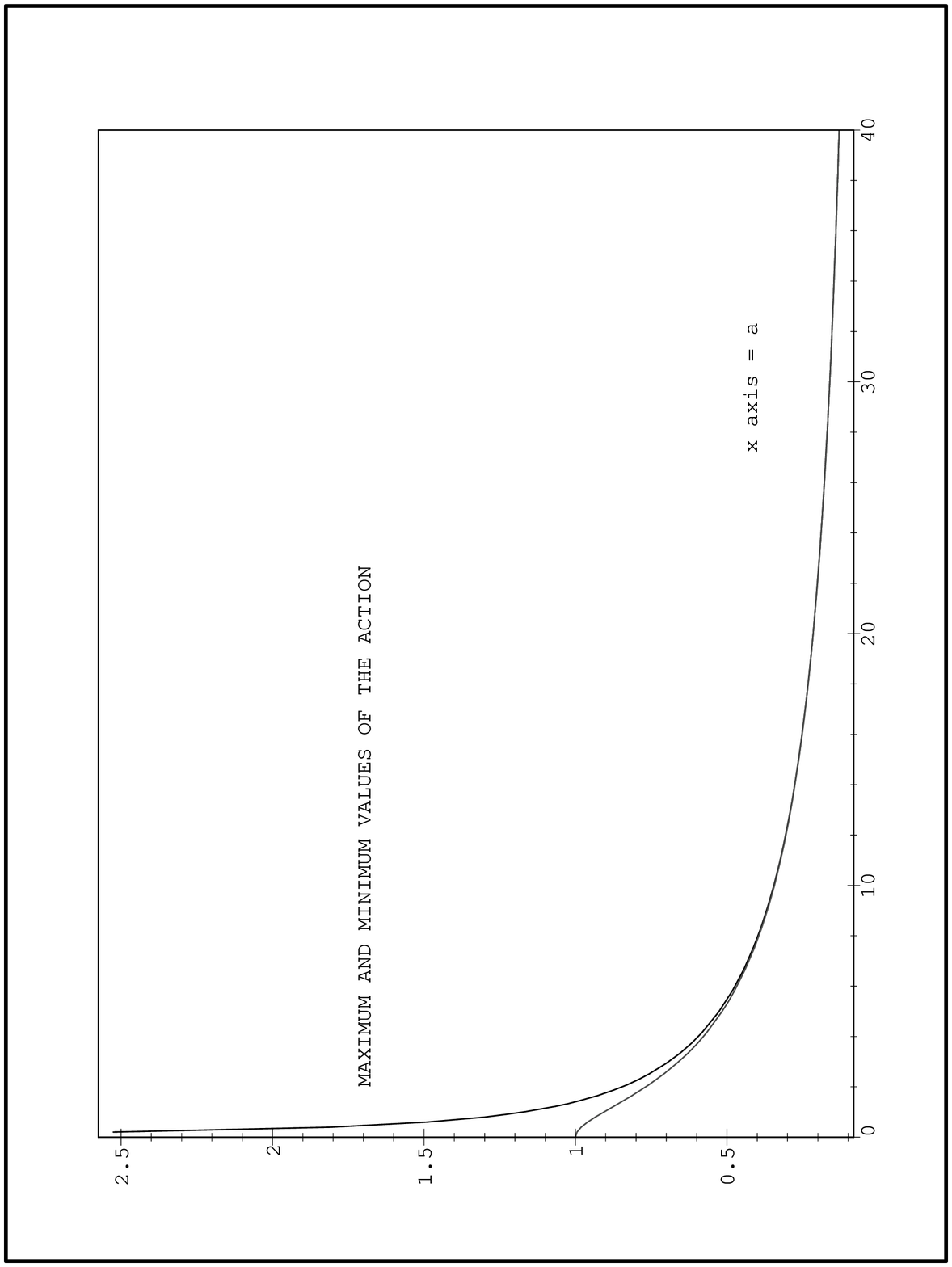}
\end{center}
\vskip 0.1 true cm
\begin{quotation}
\small
\noindent
{\bf Figure 5:}
A plot of the minimum and maximum values of the action as a function
of frequency.
\end{quotation}

At zero frequency, the minimum value of the time modulation function is
equal to 1. The maximum value becomes infinite: at small frequencies the
decay rate is completely dominated by the minimum value of the action.
Thus at zero frequency the decay rate reduces to that of
the static case, as it should. 
As the oscillation frequency increases, the maximum and minimum values
decrease monotonically and both seem to approach the 
asymptotic value zero. This
means that the (average) pair creation rate {\em increases} as the
oscillation frequency of the field increases. Finally, the action becomes
so small (and the rate so high) that the WKB approximation is no longer
valid.

Let us try to check the asymptotic behavior of the maximum and minimum. 
For the minimum, this can be done rigorously. Setting $\sin \om t_f =0$
simplifies the integral and we can identify it in terms of complete
elliptic integrals. We find
$$
   I_{\rm min} = \frac{4}{\pi} \frac{1}{a} \sqrt{\frac{1}{a^2}+1} 
 \left\{ K(\frac{a}{\sqrt{a^2 +1}}) -E(\frac{a}{\sqrt{a^2 +1}})  \right\} \ .
$$
This result was also found in \cite{MP}.
As $\om \rightarrow \infty$ ($a \rightarrow \infty$) we get
$$
   I_{\rm min} \sim \frac{\ln a}{a} \rightarrow 0 
$$
for the leading asymptotic behavior. For the maximum value, the analysis
is a bit more complicated, but we find the same leading asymptotic behavior
$$
    I_{\rm max} \sim \frac{\ln a}{a} \rightarrow 0 \ \ .
$$  

\newpage

\section{Decay of a False Vacuum}

We now turn to another problem involving tunneling with non-trivial time
dependence, whose treatment requires methods slightly different from those
we have discussed to this point.  Let us consider a scalar field
\be
  S = \int \ d^4 x \ \left[ 
    \half \pat_{\mu} \phi \pat^{\mu} \phi  - V(\phi ) \right]
\label{3a}
\ee
and take the potential to be of the form shown in Figure 6.
\begin{center}
\leavevmode
\fpsxsize 1.5in
\fpsbox{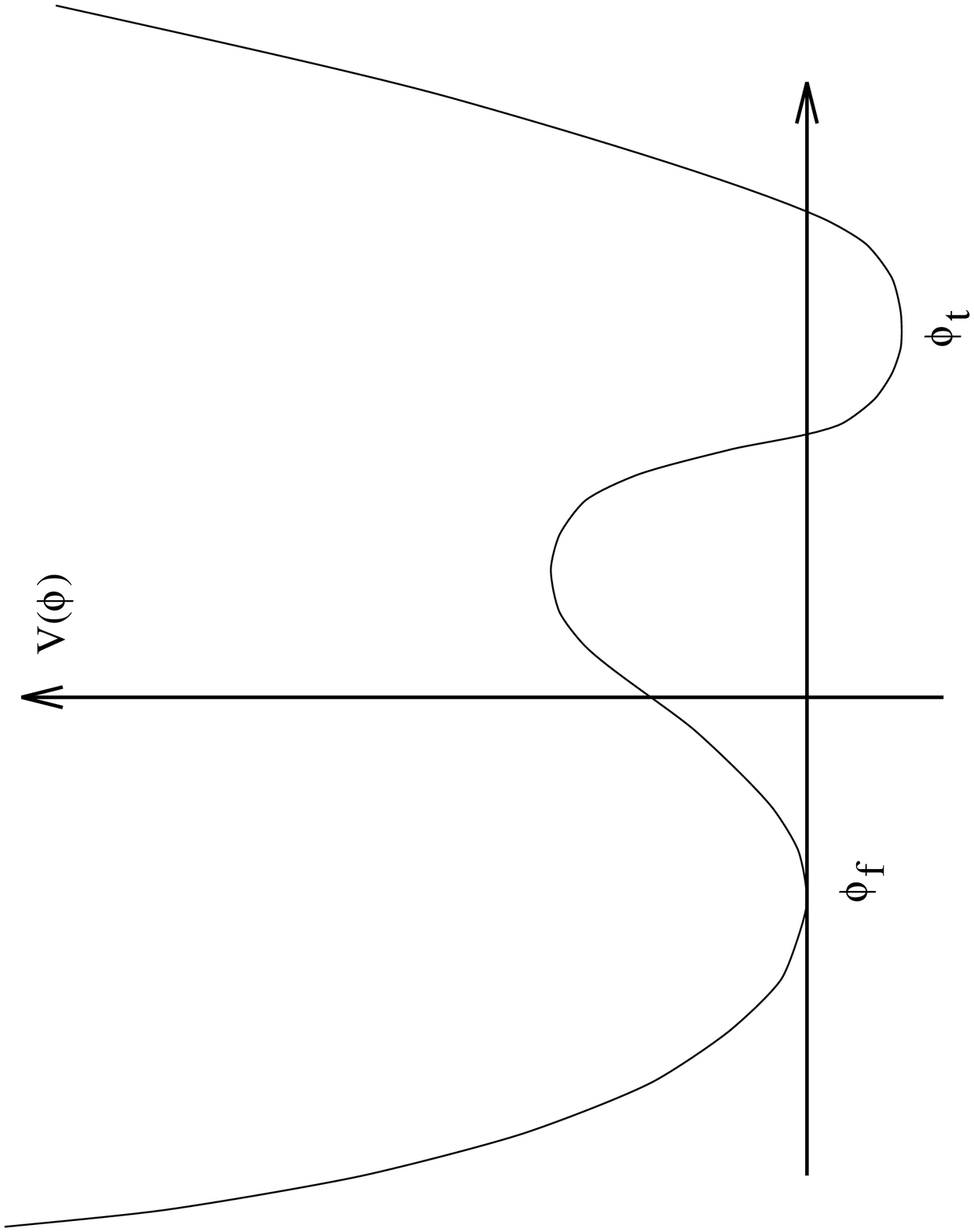}
\end{center}
\vskip 0.1 true cm
\begin{quotation}
\small
\noindent
{\bf Figure 6:}
Potential.
\end{quotation}

We take the initial state to be one in which the field is 
concentrated in the well centered at $\phi = \phi_f$. The 
field will be taken to be constant in space, but can 
have a non-trivial time dependence. In particular, we have in mind 
semi-classical looking states in which the field oscillates 
coherently in the well. Such a configuration, though stable 
classically provided the amplitude of oscillation is below 
the barrier, is expected to be unstable quantum 
mechanically. The presence of the well at $\phi  = \phi_t$ signals a decay 
process whereby the field can tunnel through the barrier. This process is 
described by bubble nucleation, meaning that regions of field concentrated 
at $\phi_t$ spontaneously form within the initial configuration and 
rapidly expand. We would like to know the rate at which bubble nucleation 
occurs, and describe the resulting bubble trajectory.  It is also important to 
determine the state of the field inside the bubble - we might 
imagine that that the
oscillations about the false vacuum outside the bubble feed 
into the interior of the
bubble, causing the field there to oscillate about the true vacuum.  
By solving the
field equations we will see that such oscillations are actually 
confined to a region
near the bubble wall, so that the interior field in the bulk of the 
bubble is frozen
at the true vacuum.  

We begin by considering the simplest case, where the field is initially 
located at the bottom of the leftmost well, $\phi (t) = \phi_f$. This, of 
course, is
the case considered by Coleman \cite{COLEMAN}. 
Our strategy will be to look for an expanding bubble solution which can be 
shrunk to zero size when evolved back along a complex time contour. 
This problem is most efficiently solved by utilizing the SO(3,1) symmetry of
the theory. However, we will later be considering initial states which 
oscillate coherently, and break the SO(3,1) symmetry to SO(3). Therefore we 
will discuss the solution in a language which explicitly uses only the 
latter symmetry.

For simplicity, we will work in the thin wall approximation, which is valid 
provided the difference in energies between the true and false vacua, 
$\rho \equiv V(\pfalse ) - V(\ptrue )$, is sufficiently 
small. The bubble solution then has the form of a spherical 
region of true vacuum separated by a thin wall from the outside 
region of false vacuum,
\be
    \phi (r,t) \approx \left\{ \begin{array}{l} \ptrue \ {\rm for} \ r<R(t) \\
                 \pfalse \ {\rm for} \ r>R(t) \ \ . \end{array} \right. 
\ee
The trajectory of the bubble wall, $R(t)$, can be determined from 
energy conservation. Consider the energy in the region $r \leq R$. 
There are two contributions to the energy. The interior of the bubble 
contributes $E({\rm inside}) = \frac{4}{3} \pi V(\ptrue ) R^3$. 
There is also an energy proportional to the area of the bubble wall 
associated with the field gradient in passing from true vacuum to false 
vacuum: $E({\rm wall})  = 4 \pi \sigma_0 R^2 / \sqrt{1- \dot{R}^2}$. 
Here, $\sigma_0$ is the energy density of the wall,
$$
    \sigma_0 = \int_{\rm wall} \ dr \ \left[\half (\phi' )^2 + V(\phi )
              \right] \ ,
$$
in the next section we will compute its value from the field equations. 
The energy $E({\rm inside})+E({\rm wall })$ must be equal to the energy present
in the region before the nucleation of the 
bubble: $E_{\rm total} = \frac{4}{3} \pi V(\pfalse ) R^3 $. So, 
\be
   \frac{4 \pi \sigma_0 R^2}{\sqrt{1 - \dot{R}^2}} 
       - \frac{4}{3} \pi \rho_0 R^3 = 0
\label{3b}
\ee
with
$$
     \rho_0 = V(\pfalse ) - V( \ptrue ) .
$$
The trajectory is then 
$$
     R(t) = \sqrt{R^2_0 + t^2} 
$$
with
\be
       R_0 = \frac{3\sigma_0 }{\rho_0} \ \ .
\label{Rcoleman}
\ee
For $t>0$ this describes an expanding bubble solution. To consider 
tunneling, we evolve the solution back to the turning point at $t=0$, and 
then try to shrink the bubble to zero size along a complex time contour. 
In the present case this step is trivial - the contour 
displayed in Figure 7. does the job.

\newpage

\begin{center}
\leavevmode
\fpsxsize 1.5in
\fpsbox{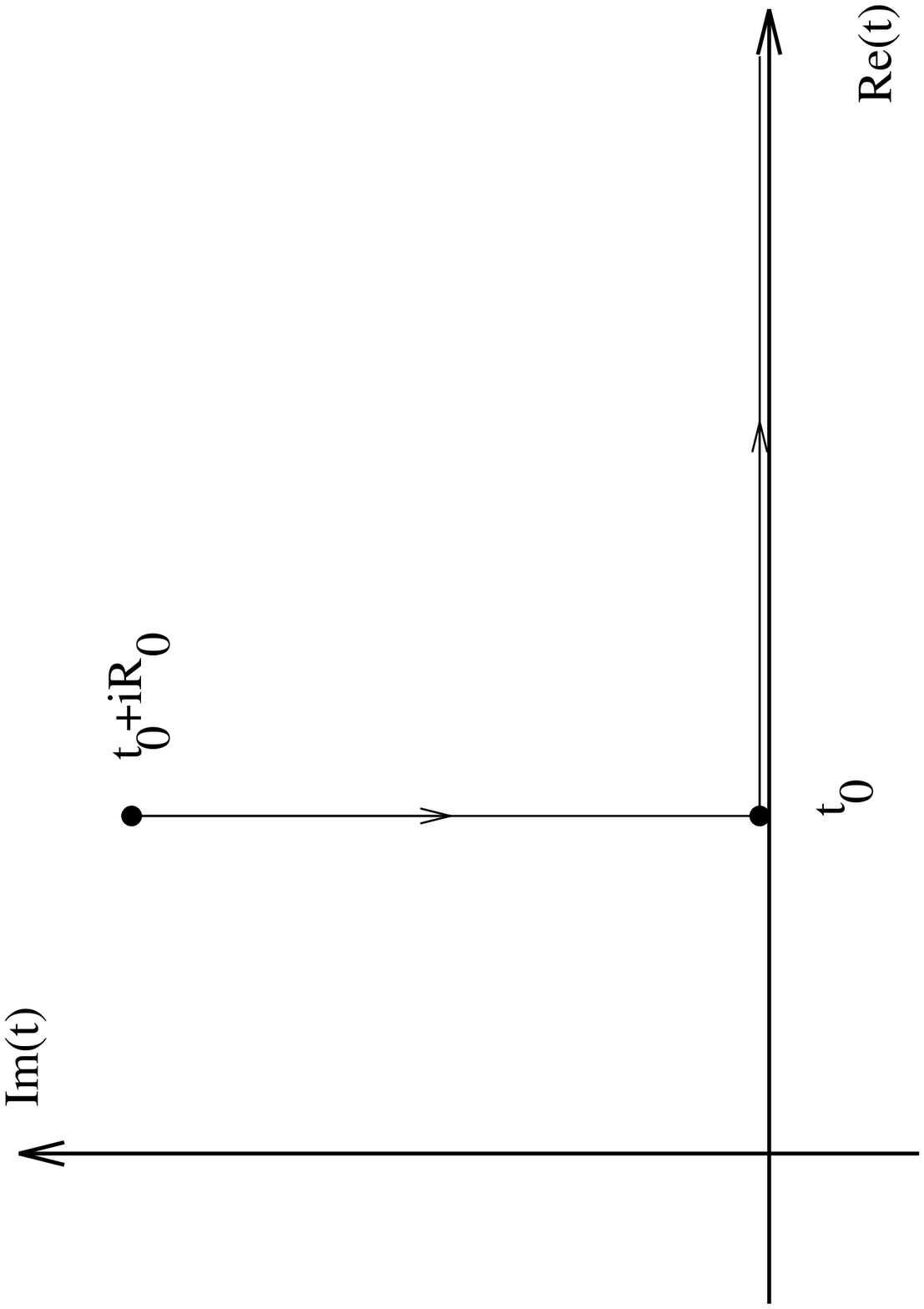}
\end{center}
\vskip 0.1 true cm
\begin{quotation}
\small
\noindent
{\bf Figure 7:}
Time path for bubble nucleation.
\end{quotation}

It remains to determine the amplitude for the tunneling process, and for 
this we require the classical action. The bubble 
has a Lagrangian
$$
     L_{\rm bubble}  =   -4 \pi \sigma_0 R^2  \sqrt{1 - \dot{R}^2}
       - \frac{4}{3} \pi V(\ptrue ) R^3 \ .
$$
Since we wish to compute the relative probability of bubble nucleation 
versus remaining in the false vacuum, what we actually want is the 
difference in action between the bubble solution and the false 
vacuum state. This is given by
$$
  S = \int \ dt \ \left[ L_{\rm bubble} 
          + \frac{4}{3} \pi V(\pfalse ) R^3 \right] 
    = - \int \ dt \ \left[ 4 \pi \sigma_0 R^2  \sqrt{1 - \dot{R}^2}
       - \frac{4}{3} \pi \rho_0 R^3 \right] \ \ .
$$
The action can be put in a useful form by 
inserting the equation of motion (\ref{3b}), and changing variables to $R$:
$$
 S = \int \ dR \ R^2 \sqrt{(\frac{4\pi}{3}\rho_0 )^2 R^2 -(4\pi \sigma_0 )^2} 
   = \int \ dR \ \frac{4\pi \rho_0 R^3}{3} 
                 \sqrt{1-\left(\frac{R_0}{R}\right)^2} \ .
$$
The action has an imaginary part coming from the part of the trajectory
$0 < R < R_0$, when the bubble is tunneling:
$$
 Im S = \frac{4\pi \rho_0}{3}
  \int^{R_0}_0 \ dR \ R^3 \sqrt{\left(\frac{R_0}{R}\right)^2 - 1} 
     = \frac{27 \pi^2 \sigma_0^4}{4 \rho_0^3} \ .
$$
The nucleation rate is then 
$$
  \Gamma \approx e^{-2 {\rm Im} [S]} 
        = \exp \left\{ -\frac{\pi^2 }{6} \rho_0 R^4_0 \right\} 
$$
which is Coleman's result.

Now let us generalize to the case where the field is initially 
oscillating around the false vacuum: $\phi = \pfalse (t)$. In  
we will study the bubble solutions with this initial condition, and we 
summarize the results here. For small 
oscillations $\pfalse (t) = \pfalse + \alpha (t)$ 
the bubble looks like 
\be
    \phi_{\rm bub} (r,t) \approx 
        \left\{ \begin{array}{l} \ptrue \; {\rm for} \ r < R-\Delta \\ 
       \ptrue + \frac{R}{r}\exp [(r-R)/\Delta] 
     \alpha (t) \; {\rm for} \ R-\Delta <r< R \\ 
    \pfalse (t) \; {\rm for} \ r > R  \end{array} \right.
\ee
where $\Delta $ is small compared to the typical size of the bubble $R$.
In other words, the field oscillations only penetrate a relatively 
small distance into the bubble; the bulk of the bubble's interior 
simply sits at the true vacuum as before. As we have already 
mentioned, although one might have expected the bubble to 
leave a state oscillating about the true vacuum, we see that 
this is not the case. 

Now we invoke the same energy considerations as before. Since the field
oscillations inside the bubble are localized near the wall, the
energy inside the bubble is essentially given by the true vacuum configuration
$\ptrue$. Thus
$$
    E_{\rm bubble} ({\rm inside }) = \frac{4}{3} \pi V(\ptrue )  R^3 \ .
$$
The bubble wall has the energy 
$$
     E_{\rm bubble} ({\rm wall }) = \frac{4\pi \sigma^{\rm bubble}_E 
                                       R^2}{\sqrt{1-\dot{R}^2}} \ ,
$$
where
$$
    \sigma_E = \int_{\rm wall} \ dr \ 
       \left\{ \half ( \dot{\phi}_{\rm bub})^2 + \half( \phi'_{\rm bub} )^2
                + V( \phi_{\rm bub} ) \right\} \ .
$$
$\sigma_E$ is time independent.  
The initial energy is also divided into two contributions: 
$$
    E_{\rm initial}({\rm inside}) = \frac{4}{3} \pi R^3 
     \ \left[\half (\dot{\phi}_f (t) )^2  + V(\pfalse (t) ) \right]  
     \equiv \frac{4}{3} \pi \rho^{\rm FV}_E R^3 
$$
and
$$
 E_{\rm initial} ({\rm wall }) = \frac{4\pi \sigma^{\rm FV}_E 
                                       R^2}{\sqrt{1-\dot{R}^2}} \ ,
$$
where
$$
 \sigma^{FV}_E~=~ \int_{\rm wall} \ dr \  \left[\half (\dot{\phi}_f (t) )^2  
                                                     + V(\pfalse (t) ) \right] 
$$
Conservation of energy then requires
$$
     \frac{4\pi \sigma_E R^2}{\sqrt{1-\dot{R}^2}}
 - \frac{4}{3} \pi \rho_E R^3 = 0 
$$
with
\bea
  \sigma_E &=& \sigma^{\rm bubble}_E - \sigma_{E}^{\rm FV} \\
 \rho_E &=& \rho^{\rm FV}_E - V(\ptrue ) \ .
\label{3b'}
\eea
We emphasize that $\rho_E$ and $\sigma_E$ are constants. This fact means
that the complex time contour relevant for tunneling runs in the purely
imaginary direction, just as in the static case. However, we expect that
this behavior is an accident of the analysis in the limit of small 
oscillations; more generally, $\rho_E$ and $\sigma_E$ will acquire time
dependence and the time contour will be a more complicated curve in the
complex time plane. Now we define
$$
      R_0 = \frac{3\sigma_E }{\rho_E}
$$
so that the trajectory is
\be
     R(t)~=~\sqrt{R_0^2+(t-t_0)^2} \ .
\label{3c}
\ee

Now that we have obtained the bubble trajectory, we turn to the 
evaluation of the decay rate.  As before, the decay rate is found by 
integrating the the action over an 
imaginary time contour running from some 
initial time $t_0$, to $t_0 +i R_0$ where
 the bubble
shrinks to zero size. The action which is to be integrated  
is the difference between the bubble action and the false 
vacuum action. The bubble action is 
$$
    S_{\rm bubble} = -\int \ dt \ \left[4\pi \sigma^{\rm bubble}_L (t) 
             R^2(t) \sqrt{1-\dot{R}^2}
                 + \frac{4\pi}{3} V(\ptrue ) R^3 \right]
$$
where
$$
     \sigma^{\rm bubble}_L = -\int_{\rm wall} \ dr \ 
\left\{ \half ( \dot{\phi}_{\rm bub})^2 - \half( \phi'_{\rm bub} )^2
                - V( \phi_{\rm bub} ) \right\} 
          = \sigma^{\rm bubble}_E - \int_{\rm wall} \ dr \ 
        \dot{\phi}^2_{\rm bub}  \ . 
$$ 
The false vacuum action is 
$$
    S_{\rm FV} = -\int \ dt \ \left[ 4\pi \sigma^{\rm FV}_L 
        R^2 \sqrt{1-\dot{R}^2}
          +  \frac{4\pi }{3} \rho^{\rm FV}_L R^3 \right]  \ ,
$$
where
\begin{eqnarray*}
     \sigma^{\rm FV}_L &=&  \sigma^{\rm FV}_E 
      - \int_{\rm wall} \ dr \ \dot{\phi}^2_f \\ 
      \rho^{\rm FV}_L &=& \rho^{\rm FV}_E - \dot{\phi}^2_f \ .
\end{eqnarray*}
The action to be integrated is thus
\be
    S = -\int \ dt \ \left[4\pi \sigma_L (t) R^2(t) \sqrt{1-\dot{R}^2}
                 - \frac{4\pi}{3} \rho_L(t) R^3 \right]
\label{bubbleaction}
\ee
where
\bea
 \sigma_L(t) &=& \sigma_E - \int_{\rm wall} \ dr  \left[ \dot{\phi}^2_{\rm bub}
              - \dot{\phi}^2_f \right] \\
 \rho_L(t) &=& \rho_E - \dot{\phi}^2_f \ \ .
\eea
A crucial point is that although $\sigma_E$ and $\rho_E$ are 
constants, $\sigma_L$
and $\rho_L$ are time dependent -- their time dependence determines the time 
dependence of the decay rate.  

The calculation has thus been reduced down to performing the 
integrals for $\sigma_L$,
$\rho_L$, and $S$.  These are straightforward to do; they can be 
done analytically
for small oscillations about the false vacuum, as is shown in Section 4.2, 
although in the general case numerical integration is required.  
The result is an expression
for the time dependent decay rate:
$$
 \Gamma (t_0) \approx \exp \{ -2{\rm Im} [S(t_0)] \} \ .
$$

\subsection{Structure of the Oscillating Bubble}

To analyze the case when the field is initially oscillating 
around the false vacuum, $\phi = \pfalse (t)$, we first 
need to determine how  
 the structure of the bubble  is altered. We will now present an 
example calculation
of the bubble solution $\phi_{\rm bub}(r,t)$ which interpolates between the
true vacuum $\phi_t$ and the oscillating initial state $\phi_f (t)$. We shall
refer to this as the oscillating bubble. 

The field equation that we need to study is
$$
     \ddot{\phi } - \frac{1}{r^2}(r^2 \phi' )' = - \frac{dV}{d\phi} \ .
$$
As an example, we consider the potential discussed by Coleman,
$$
     V(\phi ) = \frac{\lambda}{2} (\phi^2 - a^2 )^2 + 
     \frac{\epsilon}{2a} (\phi - a) \ ,
$$
where $\epsilon >0$.
The true vacuum is located at
$$
     \phi_- \approx -a - \frac{\epsilon}{8\lambda a^3} 
               + {\cal O}(\epsilon^2 ) \ , 
$$
where the potential has the 
value $V(\phi_-) \approx -\epsilon + {\cal O}(\epsilon^2)$. The false 
vacuum is located at
$$ 
   \phi_+ \approx a - \frac{\epsilon}{8\lambda a^3} 
               + {\cal O}(\epsilon^2 ) \ , 
$$
where $V(\phi_+) \approx 0 + {\cal O}(\epsilon^2 )$. 

In the standard scenario of decay from false vacuum to true vacuum, the
structure of the bubble is obtained from the static 
solution, $\phi_0(r)$, of the
field equation,
\be
      \phi''_0 + \frac{2}{r} \phi'_0 = \frac{dV}{d\phi}(\phi_0) \ .
\label{staticeqn}
\ee
$\phi_0$ interpolates between the true vacuum $\phi_-$ inside the bubble, and
the false vacuum $\phi_+$ outside the bubble, with the non-trivial
$r$-dependence  concentrated in the bubble wall. Without going into the
mathematics of the exact form of the solution, we recall that the 
qualitative behavior of the solution is as depicted in figure 8.

\newpage

\begin{center}
\leavevmode
\fpsxsize 2in
\fpsbox{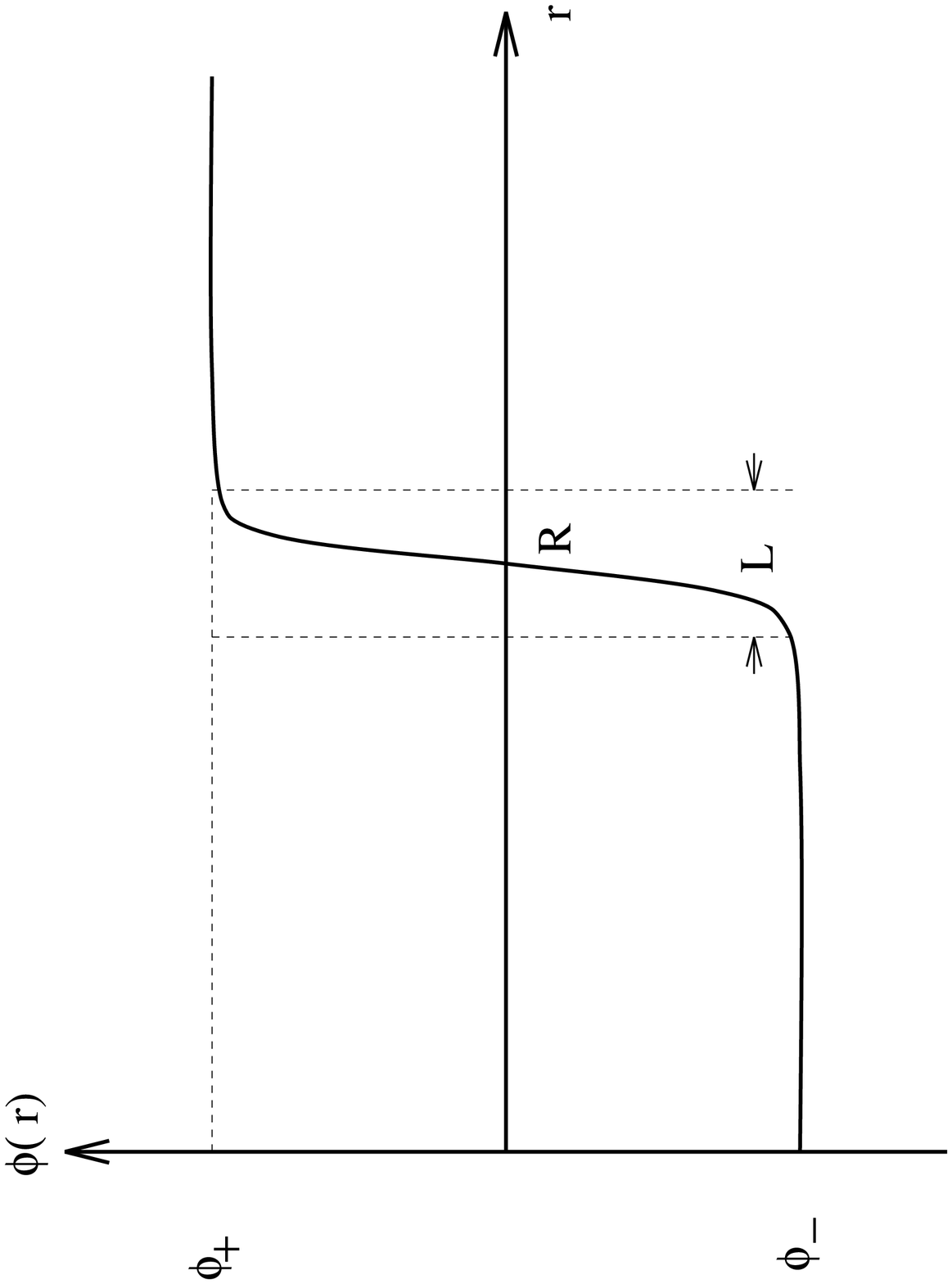}
\end{center}
\vskip 0.1 true cm
\begin{quotation}
\small
\noindent
{\bf Figure 8:}
Picture of the qualitative behavior of the static solution $\phi_0(r)$.
\end{quotation}

The radius of the bubble is $R = 3 \sigma_0 / \epsilon$ (see 
eqn. (\ref{Rcoleman})) and
the thickness $L$ of the bubble wall is of the order $L \sim 1/\mu$, where
$$
         \mu \sim \sqrt{\frac{d^2V}{d\phi^2}(\pm a) } \sim a\sqrt{\lambda} \ .
$$
An approximation to the static solution in the vicinity of the wall can
be obtained by dropping the term $(2/r)\phi'_0$ in the static field equation
(since $r \sim R \gg 0$), and dropping the small constant term $\epsilon / 2a$.
The solution of the approximate field equation is the `kink'
$$
      \phi^{\rm approx}_0 (r) = a \tanh \mu (r-R)  \ ,
$$
where $\mu = a\sqrt{\lambda}$. Its behavior is similar to that depicted in
figure 8.

Now we try to find a time-dependent solution $\phi_{\rm bub}(r,t)$, which
reduces to the coherently oscillating 
field $\phi_f(t) = \phi_+ + \alpha_0 \sin \omega t$ about the 
false vacuum as $r \rightarrow \infty$. We assume that
the amplitude $\alpha_0$ of the oscillations is small. The frequency $\omega$
of the oscillations is given by
\be
        \omega^2 = \frac{d^2 V}{d\phi^2}(\phi_+ ) = 4\lambda a^2 
        - \frac{3\epsilon}{2a^2} + {\cal O}(\epsilon^2) \ .
\label{freq}
\ee
We make the following ansatz for $\phi_{\rm bub}(r,t)$:
$$
      \phi_{\rm bub}(r,t) = \phi_0(r) + \alpha(r) \sin \omega t \ .
$$
Substituting this ansatz into the full field equation, and using the fact
that $\phi_0$ is the static solution, we obtain a linearized differential 
equation for the profile function $\alpha (r)$:
\be
     \alpha'' (r) + \frac{2}{r} \alpha' (r) + \left[ \omega^2
       - \frac{d^2V}{d\phi^2}(\phi_0) \right] \alpha (r) = 0 \ .
\label{profile}
\ee
Without knowing the exact form of the static 
solution $\phi_0$, we can still proceed
by using its known asymptotic properties. In the region outside of the bubble,
$r>R$, $\phi_0$ reduces to $\phi_+$, so $(d^2V/d\phi^2)(\phi_0)\rightarrow
(d^2V/d\phi^2)(\phi_+)$. We see that since the frequency $\omega^2$ is equal
to $(d^2V/d\phi^2)(\phi_+)$, the profile function $\alpha (r)$ must be
constant in this region. So $\phi_{\rm bub} (r,t)$ correctly reduces to
the oscillating initial configuration $\phi_f(t)$ in this region:
$$
   \phi_{\rm bub}(r,t) \rightarrow \phi_+ + \alpha_0 \sin \omega t 
  \ \ {\rm as} \ r > R \ .
$$
In the region inside the bubble, $r<R$, $\phi_0$ reduces to the true vacuum
value $\phi_-$. Thus  $(d^2V/d\phi^2)(\phi_0)\rightarrow
(d^2V/d\phi^2)(\phi_-)$. Using
$$
     \frac{d^2V}{d\phi^2} (\phi_-) = 4\lambda a^2 + \frac{3\epsilon}{2a^2}
      = \omega^2 + \frac{3\epsilon}{a^2}
$$
and denoting $k^2 \equiv 3\epsilon / a^2$, we see that the differential 
equation (\ref{profile}) reduces to a familiar equation
$$
 \alpha'' (r) + \frac{2}{r} \alpha' (r) - k^2 \alpha (r) = 0 \ .
$$
The solution for $\alpha (r)$ is,
$$
    \alpha (r) 
               = A\frac{\sinh (kr)}{kr} \ .
$$
We fix the constant $A$ by matching $\alpha (r)$ with $\alpha_0$ at
$r=R$:
$$
            A = \alpha_0 \frac{kR}{\sinh (kR)} \ .
$$
The solution for $\alpha (r)$ tells us that the 
oscillations $\alpha (r)\sin \omega t$ decay to zero inside the 
bubble, in a region of 
thickness $\Delta = 1/k$. Thus there are three scales that characterize 
the structure of the 
oscillating bubble:
\begin{enumerate}
\item The radius of the bubble $R \sim 3\sigma_0 / \epsilon$.
\item The thickness of the bubble wall $L \sim 1 / ( a \sqrt{\lambda })$.
\item The thickness of the region inside the bubble 
where the oscillations decay $\Delta \sim a / (3\sqrt{\epsilon})$.
\end{enumerate}
The relative sizes of these scales are as follows:
\begin{eqnarray*}
 \frac{\Delta}{R} &\sim& \frac{a \sqrt{\epsilon}}{\sigma_0} 
               \rightarrow 0 \ \ {\rm as} \ \epsilon \rightarrow 0 \\
 \frac{L}{\Delta} &\sim& \frac{\sqrt{\epsilon}}{a^2 \sqrt{\lambda}} 
               \rightarrow 0 \ \ {\rm as } \ \epsilon \rightarrow 0 \ .
\end{eqnarray*}
Thus, 
$$
         L \ll \Delta \ll R \ .
$$
Also, 
$$
        kR \sim \frac{\sigma_0}{\sqrt{\epsilon}} \gg 1 \ .
$$
Hence the solution for $\alpha (r)$ can be rewritten as
$$
       \alpha (r) \approx \alpha_0 \frac{R}{r} e^{(r-R)/\Delta}
$$
to a good approximation.

Finally, we would like to obtain at least an approximate 
solution for $\alpha (r)$ in the bubble wall region $r \sim R$. 
We do this by replacing $\phi_0(r)$
in the equation (\ref{profile}) with the 
approximate solution $a\tanh \mu (r-R)$, and
dropping the term $(2/r) \alpha' (r)$. Further, we approximate the frequency
(\ref{freq}) by $\omega^2 = 4\lambda a^2$. Then the equation (\ref{profile}) 
reduces to
$$
     \alpha'' (x) + \frac{6}{\cosh^2 x} \alpha (x) = 0  \ ,
$$
where $x \equiv \mu (r-R)$. This has the solution 
$$
 \alpha (r) = \frac{B}{2} [ 3 \tanh^2 (\mu (r-R)) - 1 ] \ .
$$
Since $\alpha (r) \rightarrow B$ in regions $r \stackrel{>}{<} R$ where
we know that $\alpha (r) = \alpha_0$, we set $B=\alpha_0$.

To summarize, we have found an approximate solution for the oscillating bubble:
$\phi_{\rm bub} (r,t) = \phi_0(r) + \alpha (r) \sin \omega t$, where 
$\phi_0(r)$ is the solution in the static case, modified by 
small oscillations with
a profile function $\alpha (r)$ given by
\be
     \alpha(r) = \left\{ \begin{array}{l} \alpha_0 \;
    \; \; \; \; \; \; \; \; \; \; \; \; \; \; \; \; \; \;  
    \; \; \; \; \; \; \; \; \; \; \; \; \; \; \; \; \; \; \; 
    \; \; \; \; \; \; \; r\greapp R+L/2 \\
         \frac{\alpha_0}{2} [ 3\tanh^2 ((r-R)/L) -1] \ \ \ \ \ 
      \ \ \ \ \    R-L/2 \lessapp r \lessapp R+L/2 \\
         \alpha_0 \frac{R}{r} e^{(r-R)/ \Delta} \ \ \ \ \ \ \ 
    \; \; \; \; \; \; \; \; \; \; \; \; \; \; \; \; \; \; 
    \; \; \; \; \; \; \; \; \; \; \; \; \; r \lessapp R-L/2 \ .
      \end{array} \right.
\ee
 
\subsection{Calculation of the Bubble Nucleation Rate}

Now that we have found the oscillating bubble solution $\phi_{\rm bub}(r,t)$,
we can proceed to calculate the instantaneous bubble nucleation rate 
$\Gamma (t_0) \approx \{ 2 {\rm Im}[S(t_0)] \}$, where $t_0$ is the 
time immediately
after the bubble has nucleated. 

First we need to find the quantities $\sigma_E$ and $\rho_E$ that appear
in the equation of motion of the bubble. They are:
\bea
    \sigma_E &\approx& \sigma_0 - 0.42 \ \omega^2 \alpha^2_0 L \\ \nonumber
    \rho_E  &\approx& \epsilon + \half \omega^2 \alpha^2_0  \ .
\eea
For some calculational details, see Appendix B. Since $\sigma_E,\rho_E$ are
constant, we could use energy conservation which gave us the trajectory
of the bubble (\ref{3c}):
$$
          R = \sqrt{ R^2_0 + (t-t_0)^2 } \ ,
$$
where $R_0 = 3\sigma_E / \rho_E$.

For bubble nucleation, we need a time path which shrinks the bubble to 
zero radius. Thus the time path has an imaginary segment
$$
        t = t_0 + i \sqrt{ R^2_0 - R^2 }
$$
for $R<R_0$. The above branch choice gives the correct exponential behavior
for the nucleation rate.

To calculate the imaginary part of the action, it is first convenient to 
divide the action (\ref{bubbleaction}) into two parts: $S=S_0 + S_1$, where
$$
     S_0 = - \int^{t_0}_{t_0+iR_0} \ dt \left\{ 4 \pi \sigma_E 
        R^2 \sqrt{ 1-\dot{R}^2} - \frac{4\pi}{3}\rho_E R^3 \right\} 
$$  
and
$$
   S_1 = \int^{t_0}_{t_0+iR_0} \ dt \left\{ 4\pi R^2 \sqrt{1-\dot{R}^2}
           \int^{R+L/2}_{R-L/2} dr \left( \dot{\phi}^2_{\rm bub} 
           - \dot{\phi}^2_f  \right) - \frac{4\pi}{3} R^3 \dot{\phi}^2_f 
          \right\} \ .
$$
For the first term $S_0$, the calculation proceeds as in the `static' case.
The result is 
$$
     Im S_0 = \frac{\pi^2}{12} \rho_E R^4_0 \ .
$$
For the second term, after a bit longer calculation we find
$$
     Im S_1 \approx \alpha^2_0 \pi^2 R^2_0 \left\{ \frac{(2\omega R_0)^2}{32}
         + [ 0.84 \ I_1(2\omega R_0) + \frac{1}{4} \ I_2(2\omega R_0) ]
           \cos (2\omega t_0) \right\} \ .
$$
[We have used $L\approx 2/\omega$ and dropped a negligible subleading term.]
The functions $I_n(2\omega R_0)$ are modified Bessel functions. 
The total instantaneous bubble nucleation rate is then
$$
    \Gamma (t_0) \approx
   \exp \left\{ -\frac{\pi^2}{6} \rho_E R^4_0 
        - \alpha^2_0 \pi^2 R^2_0 \left[ \frac{(2\omega R_0)^2}{16}
       + \left( 1.68 \ I_1(2\omega R_0) + \half \ I_2(2\omega R_0) 
        \right) \cos(2\omega t_0)  \right] \right\} \  \ .
$$
The decay rate is oscillatory, with the leading
correction to the static decay rate coming from the oscillatory term 
in the exponent. Recall that $R_0 \omega \sim$ radius of 
the bubble / thickness of the bubble wall, so $R_0 \omega \gg 1$.
Thus, the leading order correction is of the order
$ \alpha_0^2 R^2_0 e^{2\omega R_0 } / 2\omega R_0$: this is much larger
than what one might have anticipated. However, this is reminiscent
of what happens when a particle tunnels through an oscillating barrier. 
B\"{u}ttiker and Landauer showed \cite{BL} that the tunneling particle absorbs
quanta from the oscillating barrier; the net effect is that tunneling
becomes easier. In the leading order correction to the tunneling probability,
the amplitude of the oscillations of the barrier is multiplied by an 
exponential term, so the correction is much larger. 
As we can see in our case, small oscillations about the false vacuum 
also render the state more unstable. 

\section{Conclusion} 

In this paper we have presented a rather general approach to treating time
dependent tunneling problems, and have illustrated the method with 
some concrete 
examples.  Through the use of the WKB approximation, we found that we could 
reduce such problems to solving classical equations of motion along complex
time contours.  Even when such equations of motion are analytically 
intractable, we have shown that they are amenable to straightforward numerical
analysis. The standard approach to tunneling problems in field theory, using
instantons, is seen as a special case of the complex time method. In 
particular, the familiar procedure of evolving along the imaginary time 
direction is valid when there is no non-trivial time dependence in the problem,
but more general problems require more general complex time contours.  
   
The most straightforward and elegant application of our methods is to the
case of pair production by a time dependent electric field.  As discussed
previously, this problem has been analyzed by a  number of workers over the
years by a variety of methods; we derived a slightly more general result than
had been previously been obtained, but our main interest in the 
problem was as a
prototype for more complicated time dependent systems.  The qualitative 
features of the electric field problem carry over to these systems, the only
difference being that several steps must be performed numerically rather than
analytically.  
  
As an example of an interesting process which can be tackled by the methods
developed here, we mention again the quantum nucleation of defects in an
expanding universe.  The form of the expansion can have an important effect
on the resulting distribution of the defects.  Consider cosmic string 
nucleation, for instance.  The strings nucleate with a size equal to the
horizon, or inverse Hubble constant, which thus also sets the scale for
the duration of the complex time evolution.  If the expansion rate of the
universe varies
appreciably on this time scale, then the nucleation rate will depart from 
what a naive quasi-static analysis would indicate.  Depending on the 
cosmological model, the resulting distribution of strings is potentially
relevant.  
  
To illustrate another source of time dependence -- that arising from initial
conditions, rather than external sources -- we considered the problem of false
vacuum decay in field theory, where the initial state consists of coherent
field oscillations about the false vacuum.  Again, instanton techniques are
not directly applicable to this system. Although the analysis was rather 
involved, we were able to obtain an expression for the time dependent decay
rate in the case of small oscillations.  Of course, to see large time
dependent effects one must allow for large oscillations, but this would 
require a rather intricate computation which we have not attempted.  
But again, we stress that the
steps are, in principle, straightforward.

\bigskip

{\large \bf Appendix A}

\bigskip

In this Appendix, based on  ref. \cite{AUDRETSCH}, we 
discuss how the problem of pair production in a 
homogeneous time dependent electric field is related to  
overbarrier scattering  in nonrelativistic 
quantum mechanics; this connection is the basis
for the first quantized approach presented in Section 3.

\bigskip

We start from a relativistic field theory formulation, in which a 
charged scalar in an
electric field has the equation of motion
\be
 \{ (\pat_{\mu} + iqA_{\mu} )(\pat^{\mu} + iqA^{\mu} ) + m^2 \} \ \phi = 0 \ .
\label{App1}
\ee
In the canonical quantization approach, we seek  `natural' mode solutions
of (\ref{App1}) with suitable asymptotic properties, which we then use in 
the oscillator expansion of the field operator to identify the asymptotic
states. If we consider the background of a classical time dependent homogeneous
electric field $\vec{E} = E(t) \hat{e}_x$, it is convenient to take the
gauge choice $A^{\mu} = (A^0,\vec{A}) = (0,A(t),0,0)$, where
$A(t) = -\int \ dt E(t)$.  Then we can use the separable ansatz
$$
   \phi = \frac{1}{(2\pi )^{3/2}} \ f_{\vep }(t) \ e^{i\vep \cdot \vec{x}} 
$$
in the field equation, to reduce it to a time dependent 
generalized oscillator equation
\be
 \ddot{f}_{\vep}  \ + \ \omega_{\vep}^2 (t) \ f_{\vep} = 0 
\label{App2}
\ee
where $\omega_{\vep}^2(t) = m^2 + (p_x - eA(t))^2 + p^2_y + p^2_z
\equiv \mu^2 + (p_x - eA(t))^2$. In addition, the 
relativistic Klein-Gordon scalar
product reduces to a simple form
\be
 (f_1 ,f_2) = i(f^*_1\dot{f}_2 - \dot{f}^*_1 f_2 ) \ .
\label{App3}
\ee
 Equation (\ref{App2}) is identical to a nonrelativistic time independent
Schr\"{o}dinger equation 
$$
    \psi'' + k^2(x,E) \psi = 0 \ ,
$$
via the following identifications:
\bea
   t \leftrightarrow x \ \ &,& \ \ f_{\vep} \leftrightarrow \psi \\
  \omega^2_{\vep} (t) = \mu^2 + (p_x - qA(t))^2 &\leftrightarrow&
  k^2(x,E) = 2mE - 2mV(x) \ .
\eea
Also, the scalar product (\ref{App3}) can be seen to match with the
inner product for  wavefunctions $\psi_1,\psi_2$. If we make the more precise
identifications
\bea
    \mu^2 &\leftrightarrow& 2mE \\
    (p_x - qA(t))^2 &\leftrightarrow& -2mV(x) 
\eea
we notice that in the quantum mechanics language we must take $E>0$ and
$V\leq 0$. This suggests that we are dealing with a quantum mechanical 
overbarrier scattering problem. 

To complete this connection, we first outline the pair production calculation
in the field theory picture. If we switch the electric field on in the far
past and off in the far future, the gauge potential is asymptotically constant:
$A(t) \rightarrow A^{\rm in,out}$ as $t\rightarrow \pm \infty$. Then
the oscillator equation has two natural linearly 
independent solutions $f^{\rm in,out}$
with the asymptotic properties $f^{\rm in,out}_{\vep}(t) \rightarrow
\exp (-i\omega^{\rm in,out}_{\vep} t)$, $t\rightarrow \pm \infty$.
These give two bases for the oscillator expansion of the field operator,
corresponding to the `in' and `out' Fock spaces, as in a quantum field theory
in curved space \cite{BD}. In particular, the definition of an initial vacuum
state is different from a final vacuum state; this gives rise to particle
production. To compute the particle production rate, one must find a Bogoliubov
transformation which relates the two bases:
\be
     f^{\rm in}_{\vep} = \alpha_{\vep} f^{\rm out}_{\vep} + \beta_{\vep}
f^{\rm *out}_{\vep} \ .
\label{App4}
\ee
Then the time averaged 
pair production probability $\Gamma$ is given by 
$$\Gamma = \left|\frac{\beta_{\vep}}{\alpha_{\vep}}\right|^2 \ . $$
To see the connection to the quantum mechanical scattering problem, we
first rewrite (\ref{App4}) as 
$$
  f^{\rm out}_{\vep} + \frac{\beta_{\vep}}{\alpha_{\vep}} \ f^{\rm *out}_{\vep}
     = \frac{1}{\alpha_{\vep}} \ f^{\rm in}_{\vep} \ .
$$
Then, asymptotically as $t\rightarrow -\infty$, the l.h.s. reduces to 
$$
  e^{-i\omega^{\rm out}_{\vep}t} + \frac{\beta_{\vep}}{\alpha_{\vep}} 
  \ e^{i\omega^{\rm out}_{\vep}}
     \leftrightarrow  e^{-ikx} + R\ e^{ikx} 
$$
and the r.h.s. reduces to (as $t\rightarrow \infty$)
$$
\frac{1}{\alpha_{\vep}} e^{-i\omega^{\rm out}_{\vep}t} 
\leftrightarrow Te^{-ik'x} \ .
$$
 We see that the analysis of the
oscillator equation (\ref{App2}) in the field
theory picture exactly corresponds to an overbarrier
scattering problem in a quantum mechanical picture, with the reflection and
transmission coefficients being related to the Bogoliubov coefficients by
\begin{eqnarray*}
 |R| &=& \left| \frac{\beta_{\vep}}{\alpha_{\vep}} \right| \\
 |T| &=& \left| \frac{1}{\alpha_{\vep}} \right| \ .
\end{eqnarray*}
The time averaged pair production probability is 
$$
    \Gamma = |R|^2 \ .
$$

\bigskip

If the electric field $E(t)$ is not very rapidly varying, we can 
simplify the problem further 
with the use of WKB approximation. We use the WKB solution of the oscillator
(or Schr\"{o}dinger) equation
\be
  f \approx \exp \{-i\int  \omega dt\} \ \leftrightarrow 
     \ \exp \{-i\int k dx\} \ .
\label{App6}
\ee
The treatment of semiclassical reflection above a barrier 
requires an extension of the
WKB method \cite{WKBreview}. One method is to first find the 
complex turning points off the real axis where $\omega \sim k = 0$. 
Then, the leading contribution to the reflection coefficient 
can be found by computing
the integral $\int \omega dt$ along 
a complex time contour $C$, traveling from some initial time $t_1$ on the
real axis to the closest complex turning point $t_0$ and back (see Figure 9.).
\begin{center}
\leavevmode
\fpsxsize 1.8in
\fpsbox{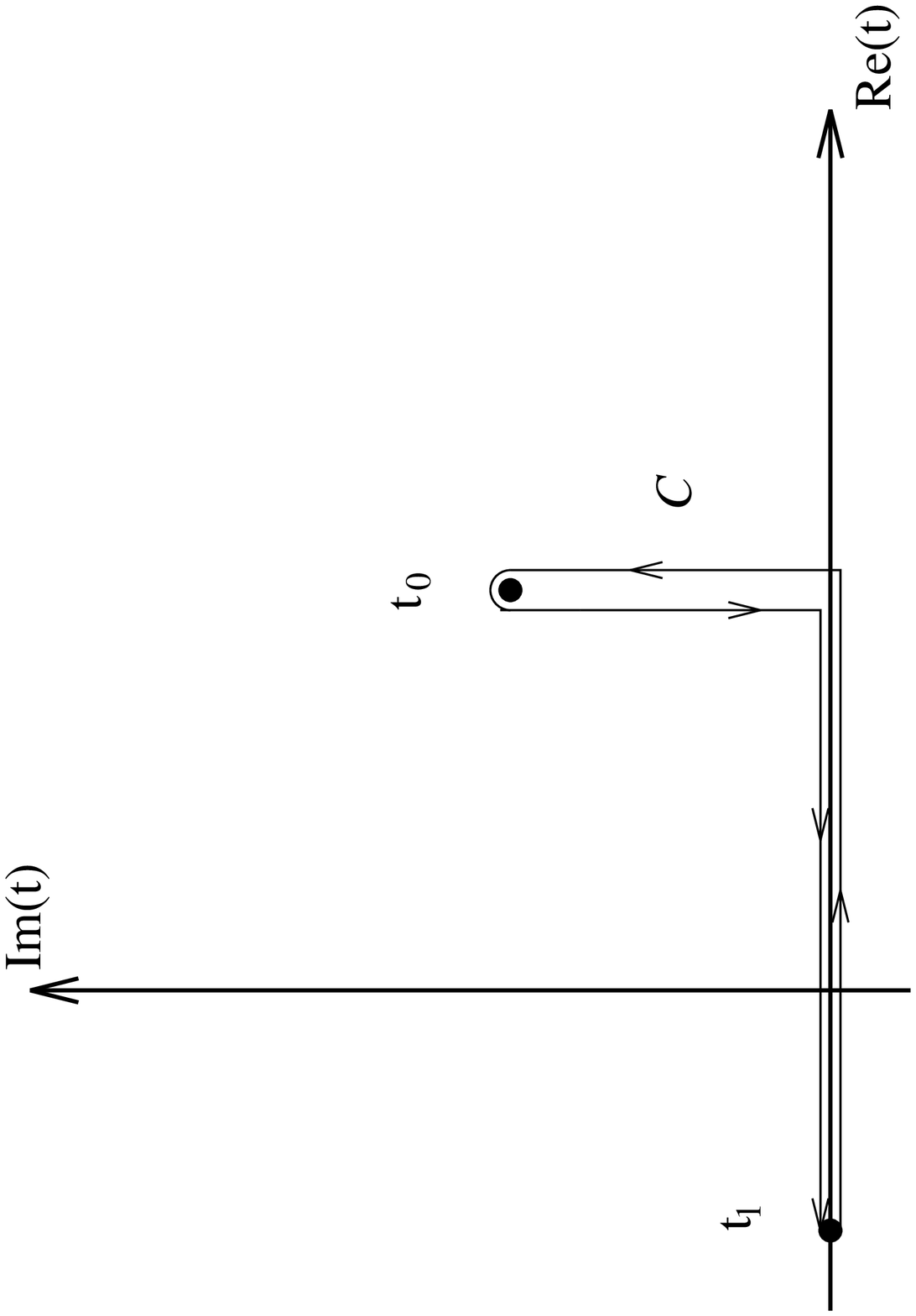}
\end{center}
\vskip 0.1 true cm
\begin{quotation}
\small
\noindent
{\bf Figure 9:}
The integration contour $C$.
\end{quotation}
This procedure yields  the reflection coefficient
\be
  |R|^2 = \exp \{ -2 {\rm Im} \ \int_C \ \omega \ dt \} \ .
\label{refcoeff}
\ee
Alternatively, we can derive the exponent of the WKB wavefunction (\ref{App6})
by starting from the action
\be
   S = - \int \ dt \ \{ m\sqrt{1-\dot{x}^2} - qA(t)\dot{x} \} 
\ee
for the relativistic charged particle in the background electric field (using
the same gauge choice as before). Using the equation of motion
$$
 v \equiv \frac{\dot{x}}{\sqrt{1-\dot{x}^2}} = -\frac{qA}{m} + p_x,
$$
and substituting $\dot{x}$, gives
\begin{eqnarray*}
S &=& -\int \ dt \sqrt{1+v^2} \\
  &=& -\int \ dt \sqrt{(p_x-qA)^2 + m^2} = - \int \ dt \ \omega \ .
\end{eqnarray*}
The WKB wavefunction (\ref{App6}) is $e^{iS}$ as expected. Therefore, 
we can use the one-particle action to calculate the reflection 
coefficient (\ref{refcoeff}); this is the basis for the 
first quantized analysis of the pair production rate presented in 
Section 3.

\newpage

{\large {\bf Appendix B}}

\bigskip

We now calculate the quantities $\sigma_E$ and $\rho_E$ that appear
in the equation of motion of the bubble. The energy density $\rho_E$
is:
\begin{eqnarray*}
\rho_E &=& \half (\dot{\phi }_f(t))^2 + V(\pfalse (t)) - V(\ptrue ) \\
       &=& \half \omega^2 \alpha^2_0 \cos^2 \omega t + V(\phi_+) 
           + \half \frac{d^2 V}{d\phi^2} (\phi_+)\ \alpha^2_0 \sin^2 \omega t +
           \epsilon \\
       &=& \half \omega^2 \alpha^2_0 + \epsilon \ .
\end{eqnarray*}
For the surface tension, we first need
$$
    \sigma^{\rm bubble}_E = \int_{\rm wall} \ dr \left\{
      \half (\dot{\phi}_{\rm bub} )^2 + \half \phi'_{\rm bub}
      + V(\phi_{\rm bub} ) \right\} \ \ .
$$
Substituting 
$\phi_{\rm bub}(r,t) = \phi_0(r) + \delta(r,t) = 
\phi_0(r) + \alpha (r) \sin \omega t$ and 
expanding to second order in $\delta$, we find
that the leading order term gives the surface tension $\sigma_0$ in the
traditional false vacuum decay,
$$
     \sigma_0 = \int_{\rm wall} \ dr \left[ \half (\phi'_0 )^2 + V(\phi_0) 
                   \right] \ .
$$
The contribution from the linear order in $\delta$ vanishes after integration
by parts and using the equation of motion for $\phi_0$. 
The contribution from the second order in $\delta$ is
$$
    \sigma_2 = \half \int_{\rm wall} \ dr \left\{ \left[ \alpha^2 
        \left(  \frac{d^2 V}{d\phi^2}(\phi_0) - \omega^2 \right) 
       + (\alpha')^2 \right] \sin^2 \omega t + \omega^2 \alpha^2 \right\} \ .
$$
Using the differential equation (\ref{profile}) for $\alpha$ (dropping
the $2\alpha' /r$ term) and 
integrating by parts, 
we find
$$
     \sigma_2 = \frac{\omega^2}{2} \int^{R+L/2}_{R-L/2} \ dr \  \alpha^2(r) \ .
$$
Substituting $\alpha (r) = \frac{\alpha_0}{2} [3 \tanh^2 ((r-R)/L) -1]$,
and combining the contributions, 
$$
        \sigma^{\rm bubble}_E \approx \sigma_0 
            + 0.16 \ \half \omega^2 \alpha^2_0 L \ .
$$
There is also the contribution from the initial state,
$$
     \sigma^{\rm FV}_E = \int_{\rm wall} \ dr \left[ \half \dot{\phi}^2_f
                     + V(\phi_f) \right] = \half \omega^2 \alpha^2_0 L \ .
$$
Then, finally, the surface tension is
$$
    \sigma_E = \sigma^{\rm bubble}_E - \sigma^{\rm FV}_E 
              \approx \sigma_0 - 0.42 \ \omega^2 \alpha^2_0 L \ .
$$


\begin{thebibliography}{123456789}

\bibitem{ERGEN} For recent work on this topic, see {\em e.g.}
K.~Ergenzinger, {\em Multiphoton Ionization as Time-Dependent Tunneling},
Z\"{u}rich U. Preprint ZU-TH-10/96 (quant-ph/9604002)

\bibitem{BI} E.~Brezin and C.~Itzykson, {\sl Phys. Rev.} {\bf D2}, 
1191 (1970)

\bibitem{MP} V.~S.~Popov, {\sl Sov. Phys. JETP} {\bf 34}, 709 (1972); 
M.~S.~Marinov and V.~S.~Popov, {\sl Fortschr. Phys.} {\bf 25}, 373 (1977) 

\bibitem{AUDRETSCH} J.~Audretsch, {\sl J. Phys. A: Math. Gen.} {\bf 12}, 1189
(1979); {\sl Gen. Rel. Grav.} {\bf 10}, 725 (1979)

\bibitem{BGV} R.~Basu, A.~H.~Guth, and A.~Vilenkin, 
{\sl Phys. Rev.} {\bf D44}, 340 (1991)

\bibitem{WIDROW} L.~M.~Widrow, {\sl Phys. Rev.} {\bf D44}, 2306 (1991)

\bibitem{SCHWINGER} J.~Schwinger, {\sl Phys. Rev.} {\bf 82}, 664 (1951);
{\bf 93}, 615 (1954)

\bibitem{strongE} W.~Greiner, B.~M\"{u}ller, 
and J.~Rafelski, {\em Quantum Electrodynamics of Strong
Fields}, Berlin; New York: Springer-Verlag (1985)

\bibitem{COLEMAN} S.~Coleman, {\sl Phys. Rev.} {\bf D15}, 2929 (1977);
C.~Callan and S.~Coleman, {\sl Phys. Rev.} {\bf D16}, 1762 (1977)

\bibitem{BL} M.~B\"{u}ttiker and R.~Landauer, {\sl Phys. Rev. Lett.} {\bf 49},
1739 (1982); {\sl Phys. Scr.} {\bf 32}, 429 (1985)

\bibitem{BD} N.~D.~Birrell and P.~C.~W.~Davies, {\em Quantum Fields in 
Curved Space}, Cambridge Univ. Press (1982)

\bibitem{WKBreview} M.~V.~Berry and K.~E.~Mount, 
{\sl Rep. Prog. Phys.} {\bf 35}, 315 (1972); L.~D.~Landau and E.~M.~Lifschitz,
{\em Quantum Mechanics. Non-Relativistic Theory}, Oxford: Pergamon (1965);
V.~L.~Pokrovskii, S.~K.~Savvinykh, and F.~R.~Ulinich, {\sl Sov. 
Phys. JETP} {\bf 34}, 879 (1958); 1119 (1958) 

\end{thebibliography}
\end{document}